\documentclass[]{aa}  
\pdfoutput=1
\usepackage{graphicx}
\usepackage{caption}
\usepackage{subcaption}
\usepackage{epsf}
\usepackage{longtable,lscape}
\usepackage{dcolumn}
\usepackage{booktabs} 
\usepackage{url}
\usepackage{float}
\usepackage{microtype}
\usepackage[varg]{txfonts}
\usepackage{natbib,twoopt} 
\usepackage{longtable}
\usepackage{lipsum}
\def\apjl{ApJ}%
\def\apjs{ApJS}%
\def\ao{Appl.~Opt.}%
\def\aap{A\&A}%
\bibpunct{(}{)}{;}{a}{}{,} 

\let\oldhat\hat

\renewcommand{\hat}[1]{\oldhat{\mathbf{#1}}}
\newcommand{{\kms}}{km\,${\rm s}^{-1}$}

\newcommand{\xmm}{{\em XMM-Newton}}

\newcommand{{\myr}}{\mbox{[$M_\odot\,{\rm yr}^{-1}$}]}

\newcommand{\msim}{\raisebox{-.4ex}{$\stackrel{>}{\scriptstyle \sim}$}}

\newcommand{{\targetname}}{HD 54879}

\newcommand{\gravunits}{\mbox{cm\,${\rm s}^{-2}$}}

\titlerunning{{\targetname}}
\authorrunning{Shenar et al.}

\begin{document}

   \title{A combined {\em HST} and {\em XMM-Newton} campaign for the
     magnetic O9.7~V star {\targetname}}

   \subtitle{Towards constraining the weak-wind problem of massive stars\thanks{Based on observations obtained with XMM-Newton, an ESA science
mission with instruments and contributions directly funded by
ESA Member States and NASA.}}

   \author{T.\ Shenar\inst{1}, 
           L.\ M.\ Oskinova\inst{1}, 
           S.~P.~J\"arvinen\inst{2},                      
           P.\ Luckas\inst{3},                
           R.\ Hainich\inst{1},      
           H.\ Todt\inst{1},                
           S.\ Hubrig\inst{2},                                 
           A.\ A.\ C.\ Sander\inst{1},            
           I.\ Ilyin\inst{2},
           W.-R.\ Hamann\inst{1}           
          }
          
   \institute{\inst{1}{Institute for physics and astronomy, University of Potsdam, 
              Karl-Liebknecht-Str. 24/25, D-14476 Potsdam, Germany}\\
              \email{shtomer@astro.physik.uni-potsdam.de}  \\    
              \inst{2}{Leibniz-Institute for astrophysics Potsdam (AIP), An der Sternwarte~16, D-14482~Potsdam, Germany} \\
              \inst{3}{International Centre for Radio Astronomy Research, The University of Western Australia, 35 Stirling Hwy Crawley, Western Australia, 6009}              
              }
   \date{Received ? / Accepted ?}


\abstract
{\object{HD 54879} (O9.7 V) is one of a dozen O-stars for which an organized atmospheric magnetic field  has been detected.
Despite their importance, little is known about the winds and evolution of magnetized massive stars.} 
{To gain insights into the 
interplay between atmospheres, winds, and magnetic fields of massive stars, we acquired UV and X-ray data of {\targetname} 
using the {\em Hubble Space Telescope} and the {\em XMM-Newton} satellite. In addition, 35 optical amateur spectra  
were secured to study the variability of {\targetname}. } 
{A multiwavelength (X-ray to optical) spectral analysis is performed using the Potsdam Wolf-Rayet (PoWR) 
model atmosphere code and the {\sc xspec} software.} 
{The photospheric parameters ($T_* = 30.5\,$kK, $\log g = 4.0\,$[\gravunits], $\log L = 4.45\,[L_\odot]$)
are typical for an O9.7~V star. The microturbulent, macroturbulent, and projected rotational velocities are 
lower than previously suggested ($\xi_\text{ph}, v_\text{mac}, v \sin i \leq 4$\,{\kms}). 
An initial mass of $16\,M_\odot$  and an age of $5\,$Myr are inferred from evolutionary tracks.
We derive a mean X-ray emitting temperature of $\log T_\text{X} = 6.7$\,[K] and an X-ray luminosity 
of $L_\text{X} = 1\cdot 10^{32}\,{\rm erg}\,{\rm s}^{-1}$.  
Short- and long-scale variability is seen in the H$\alpha$ line, but only a very long period of $P \approx 5\,$yr could be estimated. 
Assessing the circumstellar density of {\targetname} using UV spectra, 
we can roughly estimate the mass-loss rate {\targetname} would have in the
absence of a 
magnetic field as $\log \dot{M}_{B = 0} \approx -9.0\,${{\myr}}. 
The magnetic field traps 
the stellar wind up to the Alfv\'en radius $r_\text{A} \gtrsim 12\,R_*$,
implying that its true mass-loss rate  
is $\log \dot{M} \lesssim -10.2\,${\myr}.
Hence, density enhancements around magnetic stars can be exploited to 
estimate mass-loss rates of non-magnetic stars of similar spectral types, essential for resolving the weak wind problem.}
{Our study confirms that strongly magnetized stars lose little or no mass, and 
supplies important constraints on the weak-wind problem of massive main sequence stars.}


\maketitle

\section{Introduction}
\label{sec:introduction}

Massive stars ($M_\text{ini} \gtrsim 8\,M_\odot$) can outshine a million suns and radiate at energies 
that greatly exceed the ionization threshold of H, He\,{\sc i}, and He\,{\sc ii} atoms. Through their 
powerful stellar winds and their final explosion as core-collapse supernova, they provide large amounts of kinetic energy and metals to their environment.
Yet despite the importance of massive stars, our understanding of their evolution is prone to much debate. 
Along with binarity \citep[e.g.,][]{Eldridge2016, Shenar2016, Shenar2017} , rotation \citep[e.g.,][]{Georgy2012, Shenar2014, Shara2017}, and metallicity effects 
\citep[e.g.,][]{Crowther2006, Hainich2015}, 
another important question concerns the incidence of globally organized magnetic fields in massive stars and their impact on the stellar properties 
and evolution \citep[see reviews by][]{Donati2009, Langer2012}.

Roughly $5-7\%$ of the massive stars are estimated to possess
global atmospheric magnetic fields \citep{Wade2014, Scholler2017, Grunhut2017}, the majority of which are B-type. 
The first detection of an organized magnetic field in an O-type star
was reported by \citet{Donati2002} for \object{$\theta^1$\,Orionis\,C}. Since then, about ten O-stars were 
added to the sample, thanks to the Magnetism in Massive Stars \citep[MiMeS:][]{Grunhut2009, Petit2011, Alecian2014, Wade2016}, B fields in OB stars 
\citep[BOB:][]{Morel2014, Fossati2015, Hubrig2015}, and magnetic field origin \citep[MAGORI:][]{Hubrig2011} collaborations. 
It is commonly believed that massive stars may become magnetic
if they originate in significantly magnetized molecular clouds, and hence their magentic fields are fossil.
Alternatively, global magnetic fields may form as a result of merger events or dynamos produced during the pre-main sequence phase
\citep{Moss2003, Ferrario2009, Donati2009}.

Atmospheric magnetic fields can impact 
surface rotation rates via magnetic braking \citep[][]{Weber1967, ud-Doula2008}, introduce 
chemical abundance inhomogeneities and peculiarities \citep{Hunger1999}, 
and confine the stellar wind in a so-called magnetosphere \citep[e.g.,][]{Friend1984, ud-Doula2002, Townsend2005-2}. 
As a result of the latter, magnetic fields can significantly reduce 
the mass-loss from the star, favoring the creation of massive compact objects upon core collapse \citep[e.g.,][]{Petit2017}.
As the ejected matter streams along 
the field lines towards the magnetic equator, powerful collisions occur that produce 
copious X-ray emission \citep{Babel1997}.
Considering how little is known about the incidence, evolution, and impact of magnetic fields, 
studying massive magnetized stars is essential.

The subject of our study, {\targetname}, was classified as O9.7~V by \citet{Sota2011}.
\citet[][C2015 hereafter]{Castro2015} measured 
a longitudinal magnetic field reaching a maximum of $\left | B_\text{z} \right| \approx 600\,$G for {\targetname}, from which they estimated
a dipole field of $B_\text{d} \gtrsim 2\,$kG. 
The star is believed to reside in the \object{CMa OB1} association, for which an age of $\approx 3\,$Myr was estimated \citep{Claria1974}.
There are several distance estimates 
for the CMa OB1 cluster (e.g., \citealt{Claria1974}: $1.15\pm0.14\,$kpc, \citealt{Humphreys1978}: $1.32$\,kpc; \citealt[][]{Kaltcheva2000}: $0.99\pm0.05\,$kpc). 
The first data release by the {\em Gaia} satellite gives $d = 0.86_{-0.20}^{+0.36}$\,kpc for {\targetname} \citep{Gaia2016}.  
Following \citet{Gregorio-Hetem2008}, we adopt $d = 1.0\pm0.2\,$kpc in this study.

C2015 showed that a very large mass-loss rate ($\log \dot{M} \approx -5.0\,$\,{\myr}) is necessary to reproduce the H$\alpha$ emission of {\targetname}. 
This value is orders of magnitude larger than what is 
expected for late O-type dwarf stars. C2015 therefore suggested that the H$\alpha$ emission originates in the magnetosphere. In fact, late massive main sequence 
stars are generally known to exhibit mass-loss rates which are significantly lower than predicted by theory \citep{Vink2000},  
often referred to as the \emph{weak wind problem} \citep[e.g.,][]{Marcolino2009}. 

\begin{figure}[!htb]
\centering
  \includegraphics[width=\linewidth]{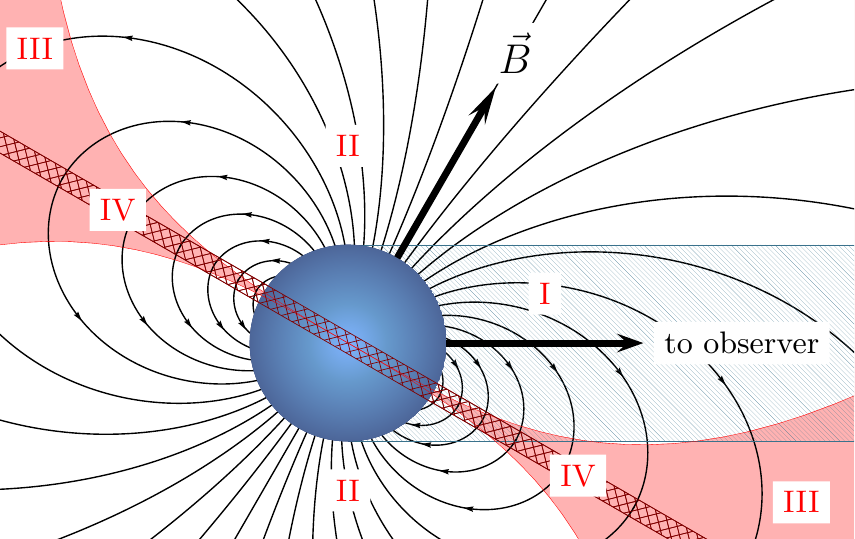}
\caption{A schematic sketch of a star with a global dipole magnetic field, illustrating the formation regions of different features 
in the spectrum of {\targetname}. 
Region I: blue-shifted UV resonance line absorption (e.g., C\,{\sc iv}, Si\,{\sc iv}, N\,{\sc v}); 
Region II: UV resonance line emission; 
Region III: shocked, X-ray emitting region; 
Region IV: recombination line emission at the magnetic equator (e.g., H$\alpha$, H$\beta$).
}
\label{fig:starsketch}
\end{figure}

Our study benefits from new UV and X-ray spectra acquired by us simultaneously
with the {\em Hubble Space Telescope} ({\em HST}), 
and the {\em XMM-Newton} satellite (see Sect.\,\ref{sec:obsdata} for details).
Complemented by optical HARPS spectra, 
the data at hand allow for a multiwavelength spectral analysis of the stellar photosphere
and magnetosphere of {\targetname} (Sect.\,\ref{sec:specan}), performed here
using the Potsdam Wolf-Rayet (PoWR) code.
The evolutionary channel of {\targetname} is discussed in Sect.\,\ref{sec:evolution}.
X-ray data are analyzed using the {\sc xspec} software (Sect.\,\ref{sec:xrays}). 
The combination of X-ray, UV, and optical data is essential, as these spectral ranges probe 
different regions of the magnetosphere and stellar wind (see Fig.\,\ref{fig:starsketch} and Sect.\,\ref{subsec:wind}). 
Additionally, we collected 35 amateur spectra covering the H$\alpha$ line to study the spectral variability of the star (Sect.\,\ref{sec:variability}).
A follow up study (J\"arvinen et al.\ in prep.) will focus on the structure and variability of the magnetic field of {\targetname}.

\section{Observational data}
\label{sec:obsdata}

We acquired three UV spectra of {\targetname} on April 30, 2016 using the {\em HST} high resolution ($R = 45\,800$) Space Telescope Imaging 
Spectrograph (STIS)\footnote{Based on observations made with the NASA/ESA Hubble Space Telescope, 
obtained at the Space Telescope Science Institute, which is operated by the 
Association of Universities for Research in Astronomy, Inc., under NASA contract NAS 5-26555. 
These observations are associated with the proposal ID 14480, PI: Hamann}. The spectra cover the spectral range $1123 - 1710\,$\AA. As no notable variability in the spectral lines 
could be identified, we co-added the three exposures to obtain a single spectrum 
with $\text{SNR}\approx 100$.  While the exposures  
are flux calibrated, they have a significant offset relative to one another of the order of a factor of two, 
likely associated with light loses originating in thermal ``breathing'' that changes the focus over orbital time scales \citep{Proffitt2017}.
The co-added spectrum was therefore recalibrated to match photometry from the TD-1 satellite \citep{Thompson1978}.

{\em XMM-Newton} observed {\targetname} on May 01, 2016 (PI: Hamann, ID: 0780180101) with an exposure time
of \mbox{$\approx40$\,ks}. All
three European Photon Imaging Cameras (EPICs: MOS1, MOS2, and PN)
were operated in the standard, full-frame mode with the medium UV filter.  
The observations were affected by 
periods of high flaring background, likely caused by soft 
protons populating Earth's 
magnetosphere\footnote{see {\it XMM-Newton Users Handbook}}. After excluding 
these periods, the useful exposure time was reduced to 29.8\,ks.
The data were reduced using the most recent calibration. The spectra and 
light-curves were extracted using standard procedures from a region with a 
diameter of about 15\arcsec.  The background area was chosen to be nearby the 
star and free of X-ray sources.

Three spectropolarimetric observations of {\targetname} were obtained with 
the HARPS polarimeter 
\citep[HARPSpol,][]{Snik2008}
attached to ESO's 3.6 m telescope (La Silla, Chile) within ESO Large programme
ID 191.D-0255 (PI Morel) on April 22, 2014 as well as on March 11 and 14, 2015. 
In this study, we make use only of the intensity spectra, which are
of high-resolution ($R = 115\,000$), high signal-to-noise ratio
(SNR$\approx 300$), and cover the wavelength range $3780 - 6912\,\AA$ with a gap at $5259 - 5337\,\AA$.

To study the spectral variability, we employed 35 spectra from the Shenton Park Observatory (SPO) taken by Paul Luckas
using a 0.35m Ritchey-Chrétien telescope equipped with a Shelyak Lhires {\sc iii} spectrograph operating at a resolution of $R \approx16\,000$ 
and producing spectra with SNR$\approx 100$. The spectra cover the spectral range 
$6500 - 6610\,\AA$ (H$\alpha$).
Spectral images were bias, dark, and flat field corrected in the normal manner, and calibrated
using Ne/Ar arc lamp spectra taken nightly and adjacent to science imaging. 
We also use seven spectra taken with the FEROS and FIES spectrographs from 
the IACOB and OWN projects \citep{Barba2010, Simondiaz2011, Simondiaz2011b, SimonDiaz2014} between the years 2009 and 2013 (see latter references 
for details). Together 
with the HARPS data, this makes a total of 45 spectra used to study the spectral variability.
The spectra were cleaned from tellurics using the 
ESO tool {\em Molecfit} \citep{Smette2015, Kausch2015} and rectified by eye.
A log containing the epochs of observation is given in Table\,\ref{tab:obslog} in Appendix\,\ref{sec:app}.

For the spectral analysis, we also use UV photometry taken by the TD-1 satellite \citep{Thompson1978}, 
$UBV$ photometry from \citet{Myers2003}, $R$ and $JHK_\text{s}$ 2MASS photometry from \citet{Zacharias2005}, and $I$-band photometry 
from \citet{Monet2003}.

\begin{figure*}[!htp]
\begin{center}
  \includegraphics[width=0.95\textwidth]{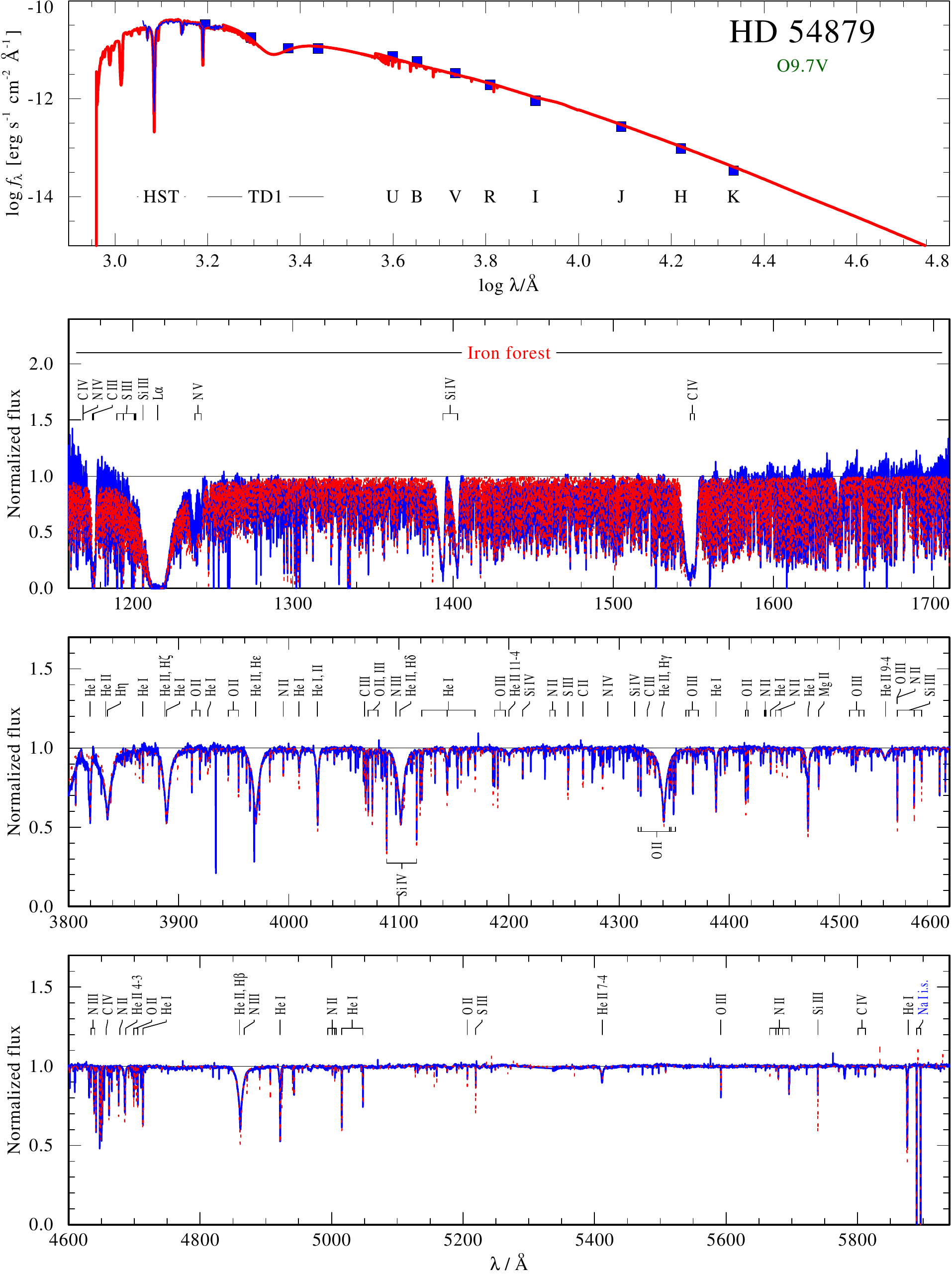}
  \caption{{\it Upper panel:} Comparison between observed photometry (blue
    squares) and the flux-calibrated {\em HST} spectrum (blue line) with the SED 
  of our best-fitting model (red line). {\it Lower panels:} Comparison between
  observed normalized {\em HST} and HARPS spectra (blue solid line) and the best-fitting model 
  (red dotted line). H$\alpha$ is shown separately (see Fig.\,\ref{fig:Balmer})}
\label{fig:master}
\end{center}
\end{figure*}

\section{Spectral analysis}
\label{sec:specan}

With the goal of inferring the fundamental stellar and wind parameters, 
we now perform a consistent, multiwavelength analysis of the optical and UV spectra at hand. 
{\targetname} was spectroscopically analyzed before (C2015),
but the analysis did not include UV data. The spectral modelling 
is performed with the PoWR model atmosphere code\footnote{PoWR models of Wolf-Rayet and OB-type stars can be downloaded at
www.astro.physik.uni-potsdam.de/PoWR}. The PoWR code solves the 
radiative transfer and statistical equations in an expanding, spherically-symmetric atmosphere, 
relaxing the assumption of local thermodynamic equilibrium (i.e., non-LTE). 
Indeed, the spherical symmetry in {\targetname} is broken by the presence of the strong magnetic dipole.
A complete non-LTE solution of a magnetized stellar atmosphere in multiple dimensions is currently unfeasible.
Nevertheless, our model provides a good approximation for the stellar photosphere, and
can deliver significant insights on the stellar wind and mass-loss, as we show in the following sections. 
For more details on the PoWR code, 
we refer to \citet{Graefener2002} and \citet{Hamann2003}.

By fitting synthetic spectra to the observations, we derive the effective temperature $T_*$, 
the surface gravity $g_*$, and the stellar luminosity $L$. 
The effective temperature $T_*$ refers to the stellar radius $R_*$, so that $L = 4\,\pi\,\sigma\,R_*^2\,T_*^4$. 
The stellar radius $R_*$ is defined at the model's inner boundary, fixed at a mean Rosseland optical depth of 
$\tau_\text{Ross} = 20$.
The velocity field consists of two regimes. In the subsonic regime, hydrostatic 
equilibrium is approached \citep{Sander2015}. In the
supersonic regime, the velocity follows the $\beta$-law with the value $\beta = 0.8$, typical for O-stars 
\citep{CAK1975, Kudritzki1989, Puls1996}. The co-moving frame radiative transfer is calculated adopting 
Gaussians for the absorption/emission coefficients with a constant Doppler width of $v_\text{Dop} = 20\,${\kms}. 
During the calculation of the emergent spectrum, $v_\text{Dop}$ is calculated via $v_\text{Dop} = (v_\text{th}^2 + \xi^2)^{1/2}$, where $v_\text{th}$ is the thermal velocity,  
and the microturbulent velocity $\xi(r)$ is assumed to grow from 
the photospheric value $\xi_\text{ph}$ to the peak value $\xi_\text{max}$ at a prespecified 
radius $R_{\xi_\text{max}}$ \citep[see detailed in][]{Shenar2015}. A depth-dependent wind clumping is assumed here, 
reaching a maximum of $D = 10$.
The synthetic profiles are convolved with Gaussians corresponding to the respective instrumental profiles.
A comparison between the best-fitting model and the observations is shown in Fig.\,\ref{fig:master}. The derived stellar parameters 
are given in Table\,\ref{tab:specan}.

\begin{table*}[!htb]
\scriptsize
\small
\caption{Derived physical parameters for {\targetname}}
\label{tab:specan}
\begin{center}
\begin{tabular}{c c c c c c c c c c c}
\hline      
\rule{0pt}{2.2ex}  
$T_*$   & $\log L$       & $\log g$                   &  $R_*$       & $M_*$       & $v \sin i$   & $\xi_\text{ph}$  & $v_\text{mac}$  & $\log \dot{M}_{B = 0}$\tablefootmark{b} & $\log \dot{M}$ & $E_{B-V}$   \\    
{[K]}   & [$L_\odot$]    & $[{\rm cm}\,{\rm s}^{-1}]$ & [$R_\odot$]  & [$M_\odot$] &   [{\kms}]   &    [{\kms}]      &       [{\kms}]  &     {\myr}                           &     {\myr} & [mag]       \\
\hline
\rule{0pt}{2.2ex}  
$30.5\pm0.5$ & $4.45\pm0.2$\tablefootmark{a} & $4.0 \pm 0.1$ & $6.1\pm1.5$\tablefootmark{a} & $14\pm7$\tablefootmark{a}  & < 4 & < 4 & < 4  & $\approx-9.0$ & $\lesssim-10.2$ & $0.35\pm0.01$  \\
\hline
\end{tabular}
\tablefoot{
Stellar parameters derived from our spectral analysis.\\
\tablefoottext{a}{The errors are dominated by the assumed uncertainty in the
  distance ($\approx20\%$, see Sect.\,\ref{sec:introduction}).} \\ 
\tablefoottext{b}{The mass-loss rate that the star would have in the absence of a magnetic field should not be confused with the true mass-loss rate.}
}
\end{center}
\end{table*}

\subsection{The stellar photosphere}
\label{subsec:photo}

\renewcommand{\arraystretch}{1.4}

\begin{figure}[!htb]
\minipage{\textwidth}
  \includegraphics[width=0.45\linewidth]{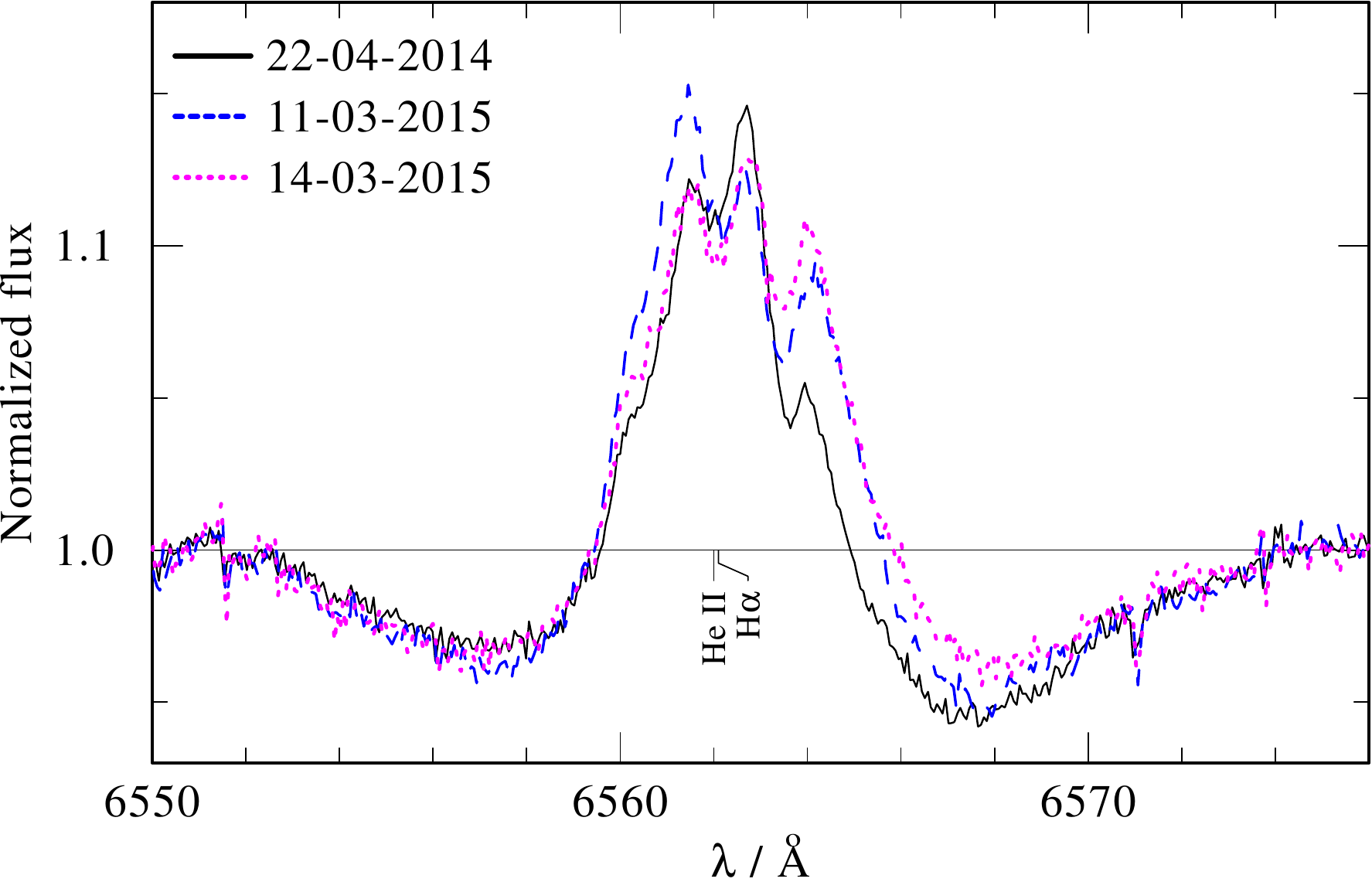}
\endminipage\hfill
\minipage{\textwidth}
  \includegraphics[width=0.45\linewidth]{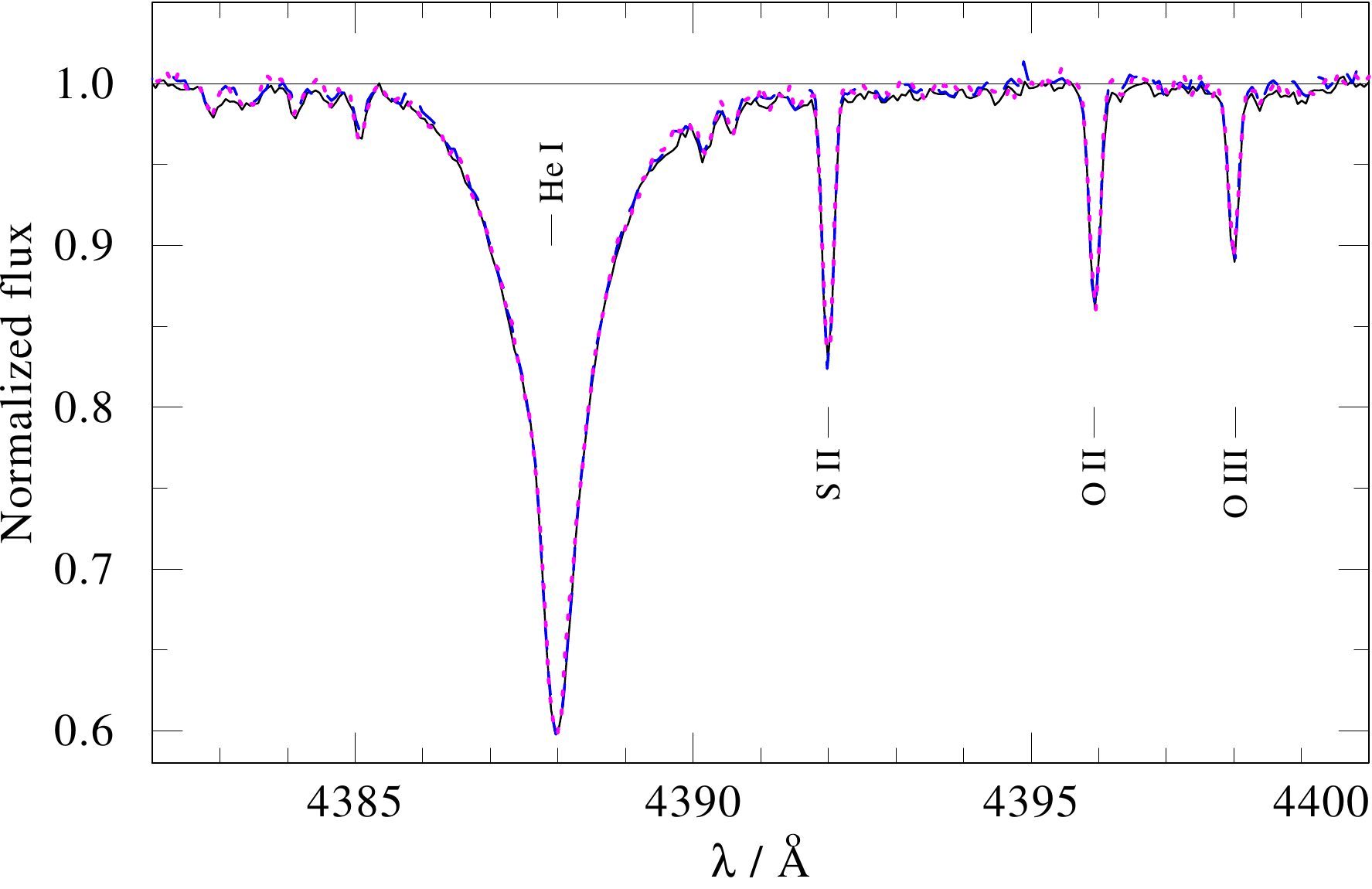}
\endminipage\hfill
\caption{The three HARPS observations focusing on H$\alpha$ (upper panel) and a few photospheric lines (lower panel).}
\label{fig:Halpha_var}
\end{figure}

Figure\,\ref{fig:Halpha_var} shows several photospheric lines
and the H$\alpha$ line as observed in the three available HARPS spectra (see Table\,\ref{tab:obslog}). As the figure illustrates, 
the H$\alpha$ line varies notably in shape, width, and 
equivalent width (EW). The intriguing three-peak profile of H$\alpha$ is persistent in all observations. In contrast, 
no variability is detected in the photospheric features in the HARPS
spectra. In fact, within the limitation set by the spectral resolution and SNR,
no notable variability of the photospheric features 
is detected in the FEROS, FIES, or HARPS spectra taken between the years 2009
and 2015.
He\,{\sc ii} lines also show no evidence for variability, 
unlike in other massive magnetic stars \citep[e.g.,][]{Grunhut2012, Hubrig2015b}. 
Assuming that the star 
changes its orientation over the span of six years because of its rotation, this
seems to suggest
that the spherical symmetry of the photospheric layers 
is virtually unbroken by the presence of the magnetic field. 
Until evidence for photospheric variability is observed in {\targetname},
a spherical symmetric model appears to be adequate for the analysis 
of its photosphere.

The surface gravity $\log g_*$ is derived primarily from the pressure broadened wings of the H$\gamma$ and H$\delta$ lines as well as He\,{\sc ii} lines in the optical spectrum 
(see Fig.\,\ref{fig:Balmer}). {H$\alpha$, and to a lesser extent H$\beta$, are strongly contaminated by emission originating 
at the magnetic equator \citep{Sundqvist2012, ud-Doula2013}. Since this is an inherently non spherically symmetric structure
(see Fig.\,\ref{fig:starsketch} and Sect.\,\ref{subsec:wind}), we do not attempt to include the disk emission in our model, 
and ignore H$\alpha$ and H$\beta$ for the estimation of $\log g_*$.
Our estimate for $\log g_*$ is found to be consistent with the value reported by C2015.

\begin{figure}[!htb]
\centering
  \includegraphics[width=\columnwidth]{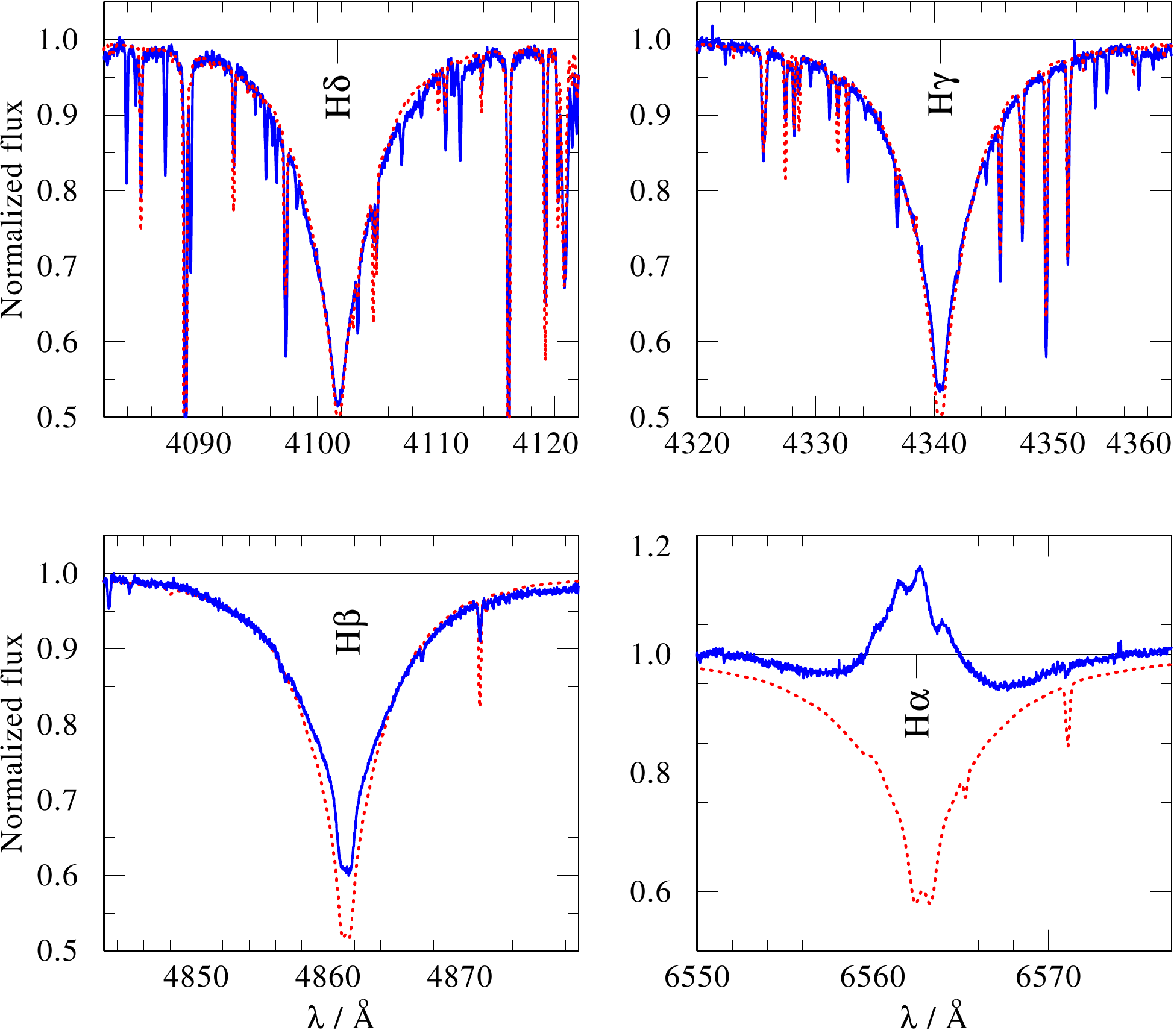}
  \caption{Our best fitting model with parameters as in Table\,\ref{tab:specan} (red dotted line) compared to the first four Balmer lines, as observed in the HARPS spectrum (blue solid line).
  The disk emission of H$\alpha$ and H$\beta$ is not included in the model.}
\label{fig:Balmer}
\end{figure}

The temperature is determined from the ratios between lines from different ionization stages of the same element,
e.g., He\,{\sc i/ii} (see Fig.\,\ref{fig:He}), N\,{\sc ii/iii/iv}, and C\,{\sc ii/iii/iv}.
The temperature we obtain ($30.5\pm0.5\,$kK) 
is significantly lower than that derived by C2015 ($33\pm1\,$kK). 
We note, however, that the temperature derived here is more consistent with our target's spectral type, O\,9.7 V, considering 
calibrations by, e.g., \citet{Martins2005}.  Moreover, a comparison of the observed photospheric spectra with TLUSTY 
model atmospheres \citep{Hubeny1995, Lanz2003} implies a similar temperature to that derived here.

\begin{figure}[!htb]
\centering
  \includegraphics[width=\columnwidth]{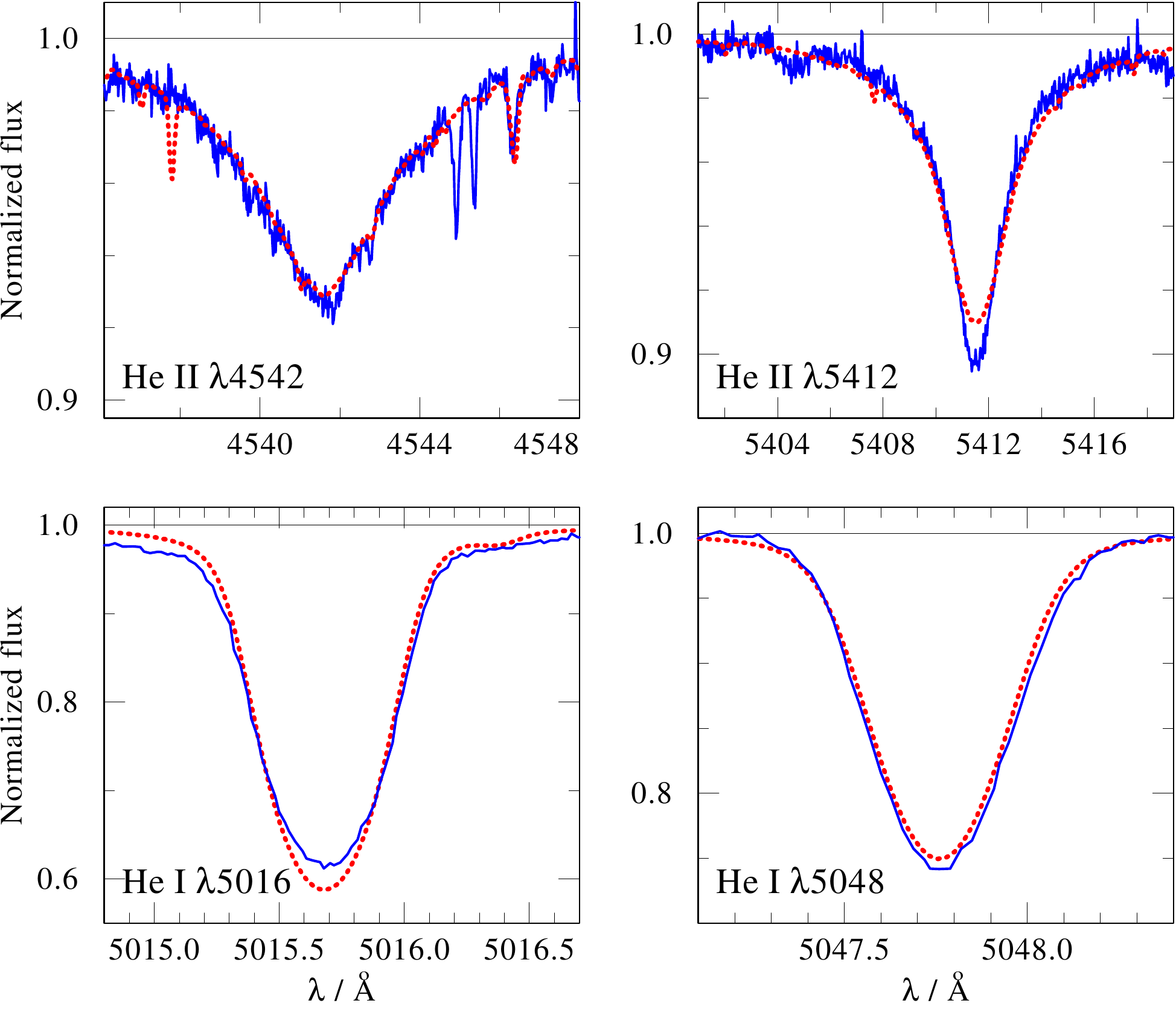}
  \caption{Same as Fig.\,\ref{fig:Balmer}, but for several He\,{\sc i, ii} lines.}
\label{fig:He}
\end{figure}

By fitting the star's spectral energy distribution (SED),  
we derive the luminosity $L$ and reddening $E_\text{B-V}$ (see upper panel of Fig.\,\ref{fig:master}). Different extinction 
laws were tested, and the SED is best reproduced using laws published by \citet{Seaton1979} and \citet{Nandy1975}. 
The luminosity derived here agrees well with C2015 when scaling $L \propto d^2$ 
to the distance used by C2015 ($d = 1.32\,$kpc), which is likely overestimated
(see discussion on the distance in Sect.\,\ref{sec:introduction}).

Finally, having constrained the fundamental parameters of the star, the abundances are inferred from the overall 
strength of spectral lines belonging to the corresponding element. Abundances that could not be derived due to the absence 
of corresponding spectral lines are assumed to have 
solar values \citep{Asplund2009}. The derived/adopted values are given in Table\,\ref{tab:abun}. We can exclude a significant 
overabundance of He compared to the solar value. Our results 
agree well with those by C2015, within errors. It is interesting to note that both C and N are found to be significantly 
subsolar, standing in contrast to reports of nitrogen enhancement in magnetic B-type stars \citep[e.g.,][]{Morel2008}.

\begin{table*}[!htb]
\scriptsize
\small
\setlength\tabcolsep{4.5pt}
\caption{Derived chemical abundances (in mass fractions) for {\targetname}}
\label{tab:abun}
\begin{center}
\begin{tabular}{l c c c c c c c c c c c c c}
\hline      
                    &  $X_\text{H}$ & $X_\text{He}$& $X_\text{C} / 10^{-3}$ & $X_\text{N} / 10^{-4}$ & $X_\text{O} / 10^{-3}$ &  $X_\text{Ne} / 10^{-3}$ & $X_\text{Mg} / 10^{-4}$ &  $X_\text{Al} / 10^{-5}$          & $X_\text{Si} / 10^{-4}$ & $X_\text{P} / 10^{-6}$ &  $X_\text{S} / 10^{-4}$  & $X_\text{Fe} / 10^{-3}$    \\    
\hline                                               
derived value       &  0.74         & 0.25         & $1\pm0.3$                & $3\pm1$                & $6\pm2$                &  $1.26$                  & $7\pm3$                 &  $4\pm2$                          &  $7\pm3$                & $5.83$                 &  $3\pm1$                 & $1.2\pm0.5$   \\    
\hline                                               
relative to solar   &  $1$          & $1$          & $0.4$                 & $0.45$                  & $1$                    &  $1$                     & $1$                     &  $0.8$                            &  $1$                    & $1$                    &  $1$                     & $1$   \\    
\hline
\end{tabular}
\tablefoot{
Values without errors were not derived and are assumed to be solar.
}
\end{center}
\end{table*}

\subsection{Rotation and turbulence}
\label{subsec:rot}

Any periodicities observed in our target are likely to correlate with its rotational period $P_\text{rot}$. It is therefore important 
to constrain the projected rotational velocity $v \sin i$ from our high resolution spectra. 
C2015 used the Fourier tool {\sc iacob-broad} \citep{SimonDiaz2014} to derive a projected rotational velocity of \mbox{$v \sin i = 7\pm2\,${\kms}}. However, 
for rotational velocities that are comparable to the microturbulent velocity, the tool strongly underestimates the involved errors, making it inapplicable for {\targetname} 
(see discussion by \citealt{SimonDiaz2014}, Sect.\ 3.4). Constraining $v \sin i$ is difficult, because its associated spectral line broadening ``competes'' with 
the microturbulent velocity $\xi_\text{ph}$, the macroturbulent velocity $v_\text{mac}$, and the thermal velocity $v_\text{th} = 2 k_\text{B} T / m$. To 
overcome this, we consider first spectral lines belonging to CNO ions. For these elements, $v_\text{th}\approx6-7\,${\kms}, which 
alone agrees with the observed line widths. Therefore, the unknowns $\xi_\text{ph}$, $v \sin i$, and $v_\text{mac}$ must be smaller than $v_\text{th}$.

\begin{figure}[!htb]
\centering
  \includegraphics[width=\columnwidth]{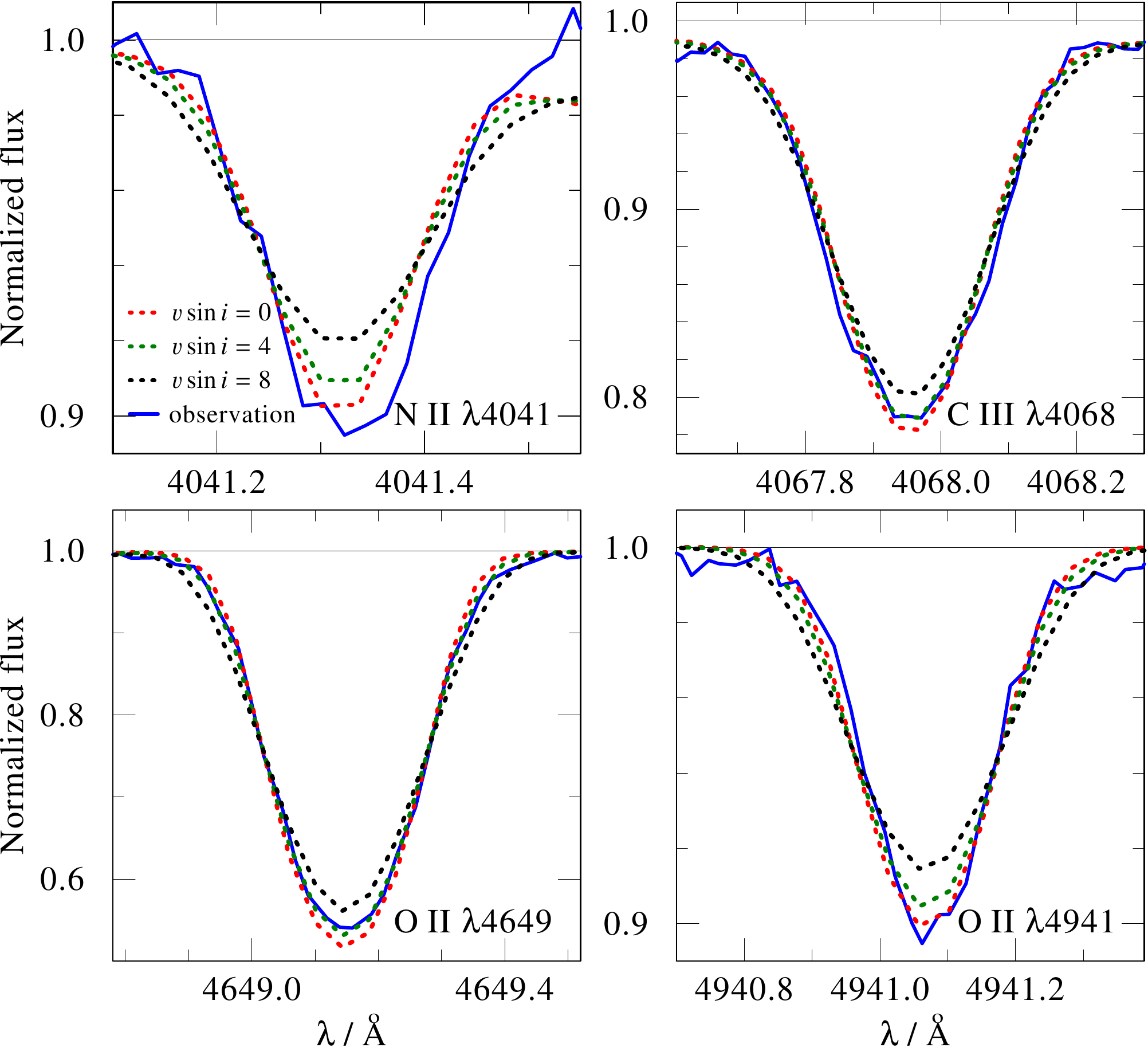}
  \caption{Selected CNO spectral lines of our best fitting model convolved with rotation profiles for $v \sin i = 0, 4$, and 8\,{\kms} (red, green, and black dotted lines, respectively), compared 
  compared to observations (blue solid line).}
\label{fig:rotation}
\end{figure} 

Figure\,\ref{fig:rotation} displays selected CNO spectral lines compared to our best fitting model convolved with rotation profiles of different $v \sin i$ values (see legend and caption)
calculated with $\xi_\text{ph} = 0\,${\kms}. 
It is evident that, for the majority of spectral lines, no broadening mechanism beyond those which are included in the calculation are required. 
Figure\,\ref{fig:rotation} shows that we can only hope to derive an upper limit for the projected rotation velocity. Based on our study, we can constrain 
this limit to $v \sin i \leq 4\,${\kms}.

The microturbulence could only be estimated approximately from the UV iron lines, which 
are subject to low thermal broadening ($v_\text{th}\approx3\,${\kms}) .
We estimate $\xi_\text{ph} \approx 3\,${\kms}, but can only reliably constrain an upper limit of $\xi_\text{ph} \leq 4\,${\kms}. 
Similarly, we can conclude $v_\text{mac} \leq 4\,${\kms}, as opposed 
to $v_\text{mac} = 8\,${\kms} reported by C2015. 

The small value of $v_\text{mac}$ derived here stands in contrast to the 
significant macroturbulent velocities reported for other magnetic O-type stars
\citep{Sundqvist2013_macro}.
A very low macroturbulent velocity 
($< 3\,${\kms}) was reported by the latter authors only for the strongly magnetized star \object{NGC\,1624-2}.
However, \citet{Sundqvist2013_macro} 
showed that this can be understood as a consequence of its extraordinary  magnetic field ($B_\text{d} = 20\,$kG), 
which is strong enough to stabilize the atmosphere at deep sub-photospheric layers, where the iron opacity peak is reached
and macroturbulence is thought 
to originate \citep[e.g.,][]{Cantiello2009}. The same argument does not hold for our target 
given its much weaker magnetic field. Rather, we suggest that the small value found here for {\targetname} is related to its low 
effective temperature (compared to the sample analyzed by \citealt{Sundqvist2013_macro}). \object{$\tau$\,Sco} is another magnetic star of 
a similar spectral type and magnetic field strength for which a low macroturbulent velocity was reported \citep{Smith1978}. Recent studies 
by \citet{SimonDiaz2017} imply that late-type massive stars exhibit systematically lower macroturbulent velocities compared to 
early-type stars, although the large scatter in $v_\text{mac}$ values prevents us from concluding this unambiguously.

\subsection{The stellar wind}
\label{subsec:wind}

Empirically derived mass-loss rates of low luminosity ($\log{L_{\rm 
bol}/L_\odot}<5.2$)  OB-dwarfs are orders of magnitude lower 
than predicted by standard mass-loss recipes 
\citep{Bouret2003, Martins2005_weakwinds,Marcolino2009,Oskinova2011, Huenemoerder2012}. This  is often
referred to in the literature as the {\em weak wind problem}.

There are various explanations as to why measured mass-loss rates 
of OB-dwarfs are so low. Some authors suggest that the wind-driving force in OB dwarfs 
is lower than predicted by current models, and that their winds are 
genuinely weak. For example, \citet{Drew1994} proposed  that the
ionization of winds  by X-rays reduces the total radiative
acceleration. 
However, \citet{Oskinova2011} showed that 
ionization by X-rays does not significantly inhibit the wind driving power 
in magnetic B-dwarfs. 
An alternative idea was proposed by \citet{Lucy2012}, who suggested 
that, in late-type O dwarfs, the shock-heating of the ambient gas 
results  in a single-component flow with a temperature of a few MK. 
This hot wind coasts to high velocities as a pure coronal wind. Hence, 
the bulk of the wind is not visible in the optical/UV, and mass-loss rates derived from these 
spectral ranges are not reliable.
High-resolution X-ray spectroscopy of O-dwarfs seem to support 
this scenario \citep[e.g.,][and references therein]{Huenemoerder2012}. 

A comparison of {\em HST} UV data of the prototypical O9.7 V star
\object{$\upsilon$ Ori} (\object{HD 36512})  with the {\em HST} data collected for our target, {\targetname}, is shown in 
Fig.\,\ref{fig:HSTobscomp}, where we focus on the two most 
prominent wind features, the resonance lines C\,{\sc iv} $\lambda \lambda 1548,1551$ and Si\,{\sc iv} $\lambda \lambda 1394, 1403$. 
It is evident 
that $\upsilon$ Ori shows no, or very little, evidence for a stellar wind. 
Assuming $\upsilon$ Ori has the same fundamental parameters as {\targetname}, and adopting a terminal velocity 
of $v_\infty = 1700\,$\kms \citep{Kudritzki2000}, a quick analysis of its UV spectrum 
using the PoWR code suggests an upper limit of $\log \dot{M} \lesssim -9.3$\,{{\myr}} for $\upsilon$ Ori.  This is lower by almost two 
orders of magnitude compared to predictions by \citet{Vink2000}, following the trend of other weak-wind stars.

\begin{figure}[!htb]
\centering
  \includegraphics[width=\columnwidth]{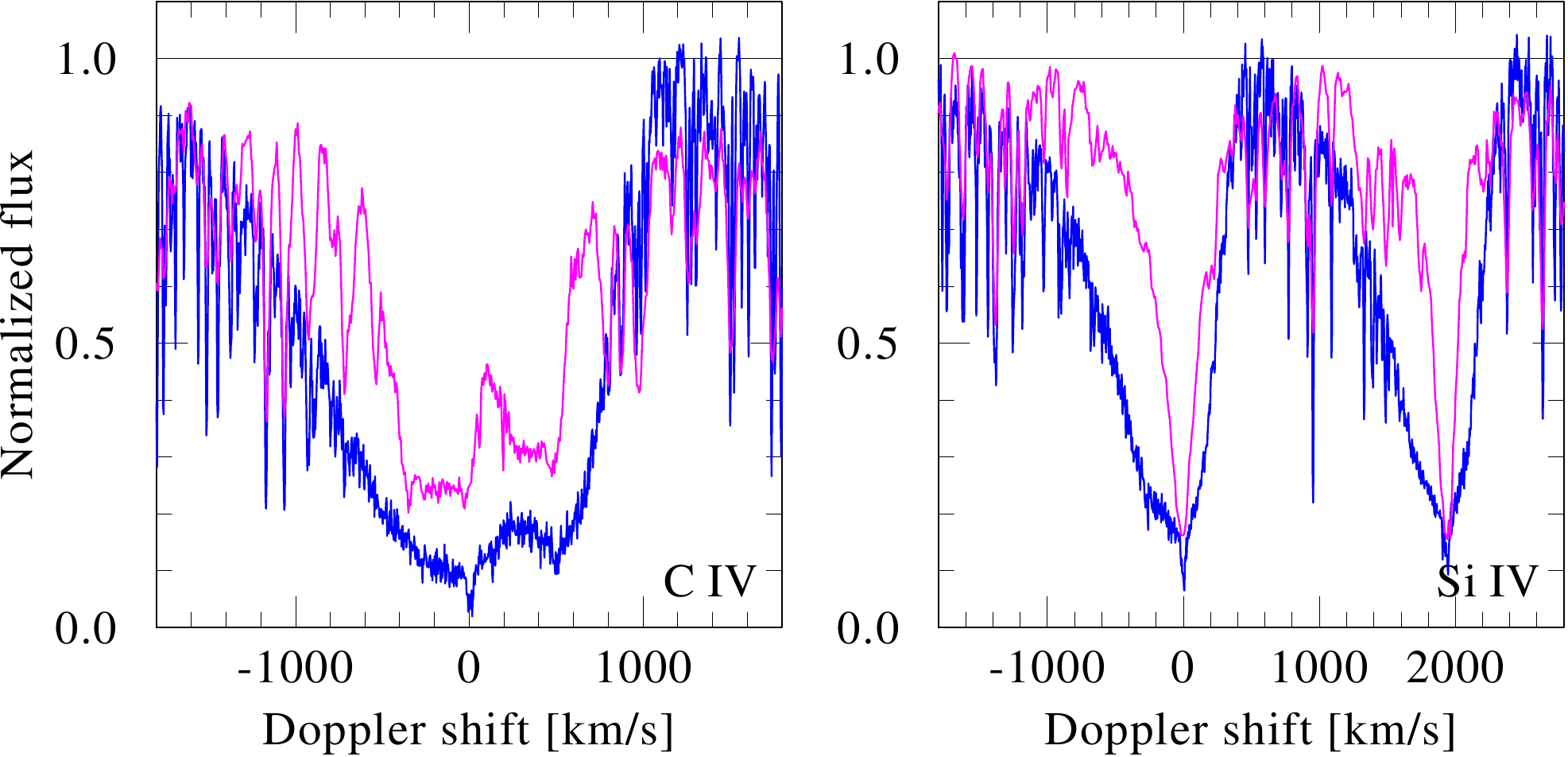}
  \caption{Comparison of normalized {\em HST} observations 
  of the prototypical O9.7 V star HD\,36512 (ID: 13346, PI: Ayres, pink solid line) and of {\targetname} (blue solid line) in Doppler space. Shown are the 
  resonance lines C\,{\sc iv} $\lambda \lambda 1548, 1551$ (left panel) and Si\,{\sc iv} $\lambda \lambda 1394, 1403$ (right panel).}
\label{fig:HSTobscomp}
\end{figure} 

In contrast to $\upsilon$ Ori, our target shows a clear asymmetry that is suggestive 
of absorption stemming from matter surrounding the star. 
At first glance, this seems to suggest that the magnetic field 
in fact spurs mass-loss from the star, contrary to expectation. However, 
the profiles of the resonance lines of {\targetname} have unusually shallow blueshifted edges,  
suggesting the presence of a large velocity dispersion, reaching a maximum speed of $\approx1000\,${\kms}. 
Similar line profiles were reported in other studies of magnetic stars \citep[e.g.,][]{Marcolino2013, Naze2015}.

The magnetic field in {\targetname} dominates the behavior of the stellar wind for radii smaller
than the Alfv\'en radius $r_\text{A}$, which is estimated to $r_\text{A} \gtrsim 12\,R_*$ in this work (see below). 
To understand the line profiles of the Si\,{\sc iv} and C\,{\sc iv} resonance lines (Fig.\,\ref{fig:HSTobscomp}), as well as the 
prominent H$\alpha$ emission (Fig.\,\ref{fig:Balmer}), we refer the reader to Fig.\,\ref{fig:starsketch}.
The absorption of the UV resonance lines 
is mostly blueshifted, because it is formed in front of the stellar disk, where the material moves mostly towards the observer 
(region I). Some emission in the resonance lines (blueshifted and redshifted) is formed in the less dense region II, but 
it competes with line absorption originating in the stellar photosphere. Moreover, X-rays formed in 
region III \citep{Owocki2016} ionize the Si\,{\sc iv} and C\,{\sc iv} ions and reduce their emissivity/opacity. 
Finally, as a recombination line, the emission in H$\alpha$ scales with 
$\rho^2$, $\rho$ being the density. Therefore, $H\alpha$ emission stems primarily from 
the magnetic equator  (region IV), where $\rho$ is orders of magnitudes
larger. In case of a pronounced shock retreat, the formation of a disk-like structure may be inhibited
\citep{ud-Doula2014}. However, 
H$\alpha$ is still expected to trace dense environments that
strongly deviate from spherical symmetry, and is therefore not modelled in the framework of this study.
Both the dynamic nature of the disk as well as possible co-rotation contribute to the width of the H$\alpha$ feature.
The presence of H$\alpha$ emission or broadened, asymmetric lines in the UV
is therefore not related to the stellar wind, but to the magnetosphere, and they do not imply 
a mass-loss from {\targetname} \citep[see also][]{Naze2015, Erba2017}.

To illustrate the inability of a standard, spherically-symmetric outflow to reproduce the observations, we plot in Fig.\,\ref{fig:HSTmodcomp} 
several synthetic spectra calculated with the parameters given in Table\,\ref{tab:specan}, but with $v_\infty = 900$\,\kms (roughly corresponding to the 
observed blue edge of the lines), a standard 
wind turbulent velocity ($0.1\,v_\infty$), and different mass-loss rates. It is evident from the figure that no combination of wind parameters
can reproduce the observed shallow profiles, which indicates that the stellar wind is heavily affected by the 
magnetic field.

\begin{figure}[!htb]
\centering
  \includegraphics[width=\columnwidth]{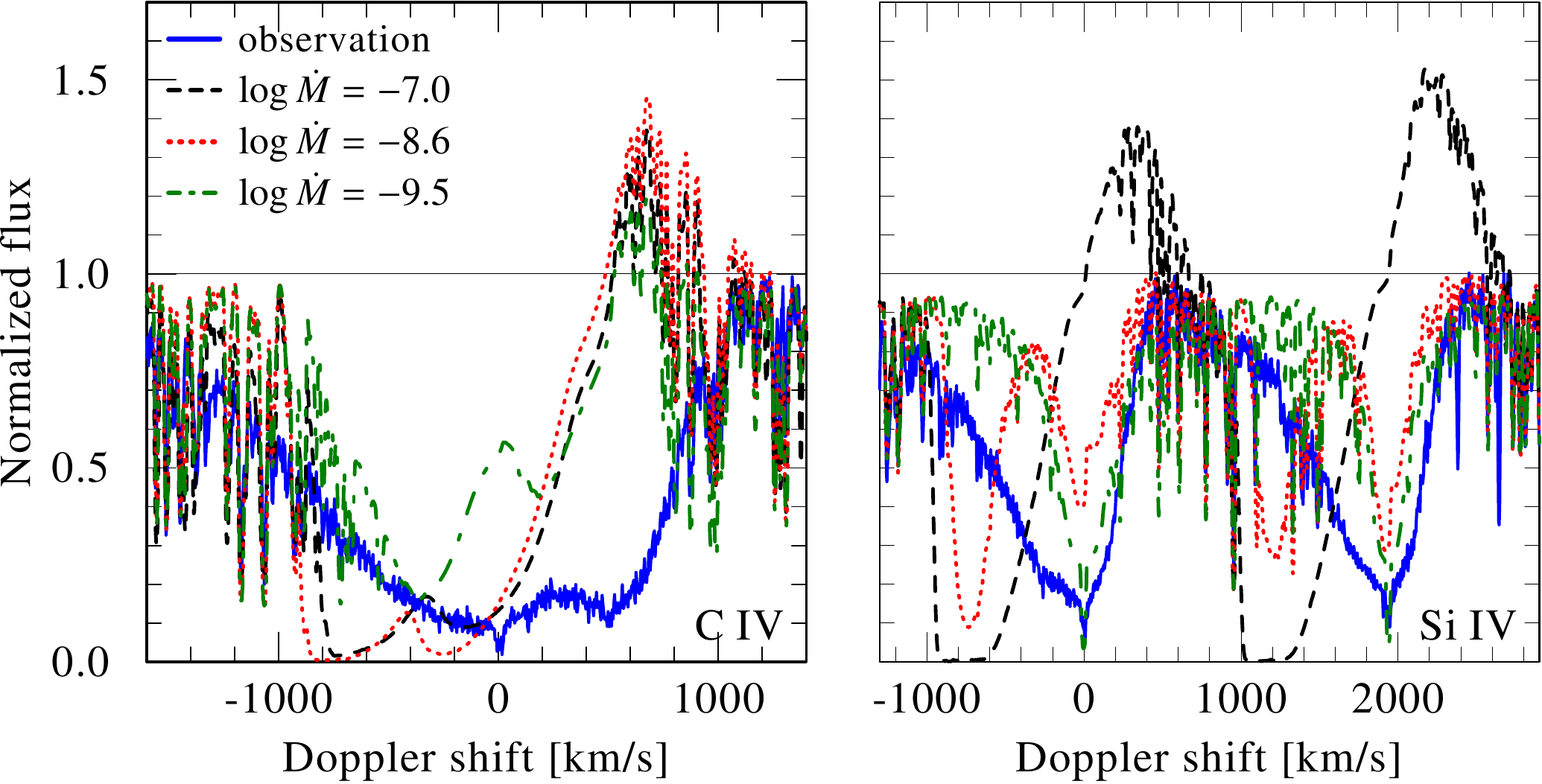}
  \caption{Normalized {\em HST} observations of {\targetname} (blue solid line) compared to synthetic spectra with 
  a terminal velocity of $v_\infty = 900\,${\kms} and different mass-loss rates (given in the legend in ${\myr}$).}
\label{fig:HSTmodcomp}
\end{figure} 

Solving the full non-LTE radiative transfer in the presence of a magnetic field in 3D is currently not feasible and beyond the scope of this 
study. Instead, we simulate the motion of the matter along the field lines by a turbulent velocity which is strongly enhanced 
close above the stellar surface, at $r \approx 1.1\,R_*$. We find the best fit for the combination of $v_{\infty, {\rm sph}} = 300\,${\kms}, 
$\xi_\text{wind, sph} = 500\,${\kms}, and $\log \dot{M}_\text{sph} = -8.8\,{\myr}$. X-rays ionization is approximately accounted 
for in a spherically-symmetric fashion (see Sect.\,\ref{subsec:xraywind}).
The results are shown in Fig.\,\ref{fig:HSTfinal}. 
Note that these parameters do not correspond to actual physical parameters, since 
they are obtained via a spherically symmetric model which assumes a constant outflow. Rather, these parameters serve to 
reproduce the conditions in the formation region of the resonance lines. For example, the model 
predicts that the density at the main line forming region, at $r \approx 1.1-2\,R_*$, is 
$\rho \approx 10^{-15}- 10^{-16}\,{\rm g}\,{\rm cm}^{-3}$.

\begin{figure}[!htb]
\centering
  \includegraphics[width=\columnwidth]{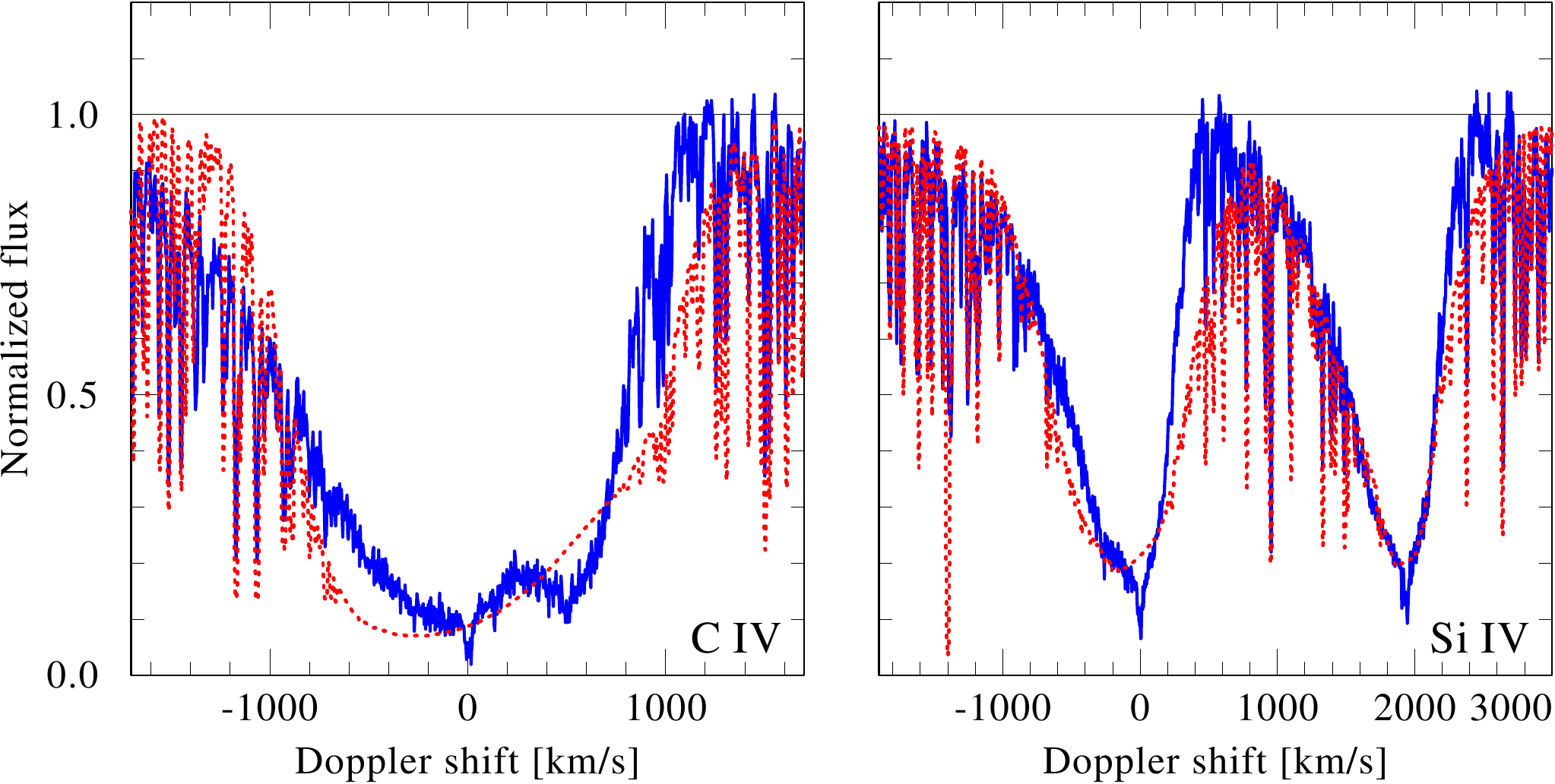}
  \caption{Normalized {\em HST} observations of {\targetname} (blue solid line) compared to our best-fitting model, which accounts for 
  a large microturbulence and superionization by X-rays.}
\label{fig:HSTfinal}
\end{figure} 

By comparing the derived densities to predictions by an 
analytical model derived for stars with a constant outflow and a global dipole magnetic field \citep{Owocki2016}, we can give a rough 
estimate to the mass-loss rate {\targetname} would have in the absence of a magnetic field, $\dot{M}_{B = 0}$.
This model assumes three components: (a) an upflow along
  the dipole lines towards the loop apex, (b) a shocked, X-ray emitting plasma, 
  and (c) a cooled downflow from the dipole apices back towards the star. Assuming
  simplistically a $\beta$-type law for the expansion speed, the model predicts the
  densities $\rho_\text{u}$, $\rho_\text{s}$, and $\rho_\text{d}$  (Eqs.\ 10,
  20, and 25 in \citealt{Owocki2016}) as a
  function of the polar coordinates $(r, \theta)$ of the
  upflowing, shocked, and downflowing plasma as a function of $v_\infty$,
  $R_*$, $M_*$,  and $\dot{M}_{B = 0}$.\footnote{The model also introduces the
    parameter $\delta$ which describes the location of onset of matter infall
    along the dipole loop. The impact of this parameter is, for our
    estimation, negligible.} Since
  the P-Cygni absorption originates in the upflowing and downflowing plasma in
  the magnetosphere, we compare the densities derived from our spherical
  models with the sum $\rho_\text{u} + \rho_\text{d}$. 
  Evaluating this for a typical viewing angle of $\theta = 60^\circ$ at 
different radii between $1.1$ and $2.0\,R_*$, and assuming $v_\infty =
1700\,${\kms} and the parameters derived in our model, we obtain
values for $\log \dot{M}_{B = 0}$ that range between $-9.2$ and
$-8.8\,${{\myr}}. This is smaller than the values predicted by
\citet{Vink2000} ($\approx -7.7$\,\myr) by more than an order of magnitude,
but in line with 
values reported for other stars of similar spectral type \citep[e.g.,][]{Marcolino2009}. 

Since our radiative transfer assumes spherical symmetry, our result should be considered  
as an order-of-magnitude estimate ($\pm 0.5\,$dex), and not as an accurate derivation of  $\log
\dot{M}_{B = 0}$. From $\log \dot{M}_{B = 0} \approx -9.0\,${\myr} and 
$B_\text{d} \gtrsim 2\,$kG, and assuming again $v_\infty = 1700$\,\kms, we can
constrain the Alfv\'en radius to $r_\text{A} \gtrsim 12\,R_*$. 
From this, we can then estimate the true mass-loss from the star to $\log \dot{M} \lesssim -10.2\,${\myr}  \citep[see 
equations 1-3 of][]{Petit2017}.
However, we cannot exclude the possibility that some mass is lost from the star as a hot, X-ray emitting wind.

Importantly, while the magnetic field suppresses mass-loss from the star, it causes a density enhancement around it that leaves a spectroscopic signature.
Therefore, the analysis of massive magnetic stars using appropriate models may enable one, in principle, to predict 
the mass-loss rates of non-magnetic stars of a similar spectral type, which cannot be measured otherwise.
Magnetic stars can therefore help to resolve the weak-wind problem.

\subsection{The effect of X-rays on the stellar wind}
\label{subsec:xraywind}

The X-rays present in the stellar atmosphere are expected to affect the ionization balance via 
K-shell Auger ionization \citep{Auger1923}. This effect is known to lead to high ionization stages 
such as N\,{\sc v} and O\,{\sc vi} \citep{Cassinelli1979, Oskinova2011}. Moreover, X-rays 
affect diagnostic wind lines such as the resonance lines of C\,{\sc iv} and Si\,{\sc iv}. In our case, 
the N\,{\sc v} $\lambda \lambda 1239, 1243$ resonance lines are clearly
present in the {\em HST} data, while our model predicts that N\,{\sc v} 
is not significant in the stellar wind when not including X-rays. This suggests that X-rays contribute to the 
appearance of the UV spectrum.

Auger ionization via X-rays is accounted for in our model by 
assuming optically thin, thermally emitting filaments of shocked plasma embedded in the wind \citep{Baum1992}. While the topology of 
the X-ray emitting 
plasma may be more complex in reality, the important quantity here is merely the amount of ionizing radiation. 
The observed X-ray spectrum is characterized by two temperature components $T_\text{X,1}$ and 
$T_\text{X,2}$ and corresponding filling factors $X_\text{fill, 1}$ and $X_\text{fill, 2}$. 
These parameters are adopted from our analysis presented in Sect.\,\ref{sec:xrays}, where only the two 
dominant components are accounted for.
The onset radius of these filaments is 
set to $R_0 = 1.05\,R_*$. Slightly different values deliver similar results, but onset radii which are too large trivially do not affect 
the spectra. 

In Fig.\,\ref{fig:HSTXrays}, we compare the {\em HST} observations of the N\,{\sc v}\,$\lambda \lambda 1239, 1243$ and Si\,{\sc iv}\,$\lambda \lambda 1394, 1403$
resonance lines to our best-fitting model with and without X-rays. As discussed above, the N\,{\sc v} resonance doublet is clearly seen in the observations but 
not in the model without X-rays. With the inclusion of X-rays, the N\,{\sc v} lines appear. The N\,{\sc v} line profiles have a similar shape to the C\,{\sc iv} and 
Si\,{\sc iv} resonance lines, suggesting that they too form in the magnetosphere. The strength of the lines implies a nitrogen abundance slightly lower than derived 
from our photospheric analysis by a factor of $\approx 0.7$, but this is within our given uncertainty. The X-rays also significantly influence 
the resonance lines of Si\,{\sc iv} (see Fig.\,\ref{fig:HSTXrays}, right panel) and C\,{\sc iv}. Including X-rays, our model successfully reproduces 
the main characteristics of our observations.

\begin{figure}[!htb]
\centering
  \includegraphics[width=\columnwidth]{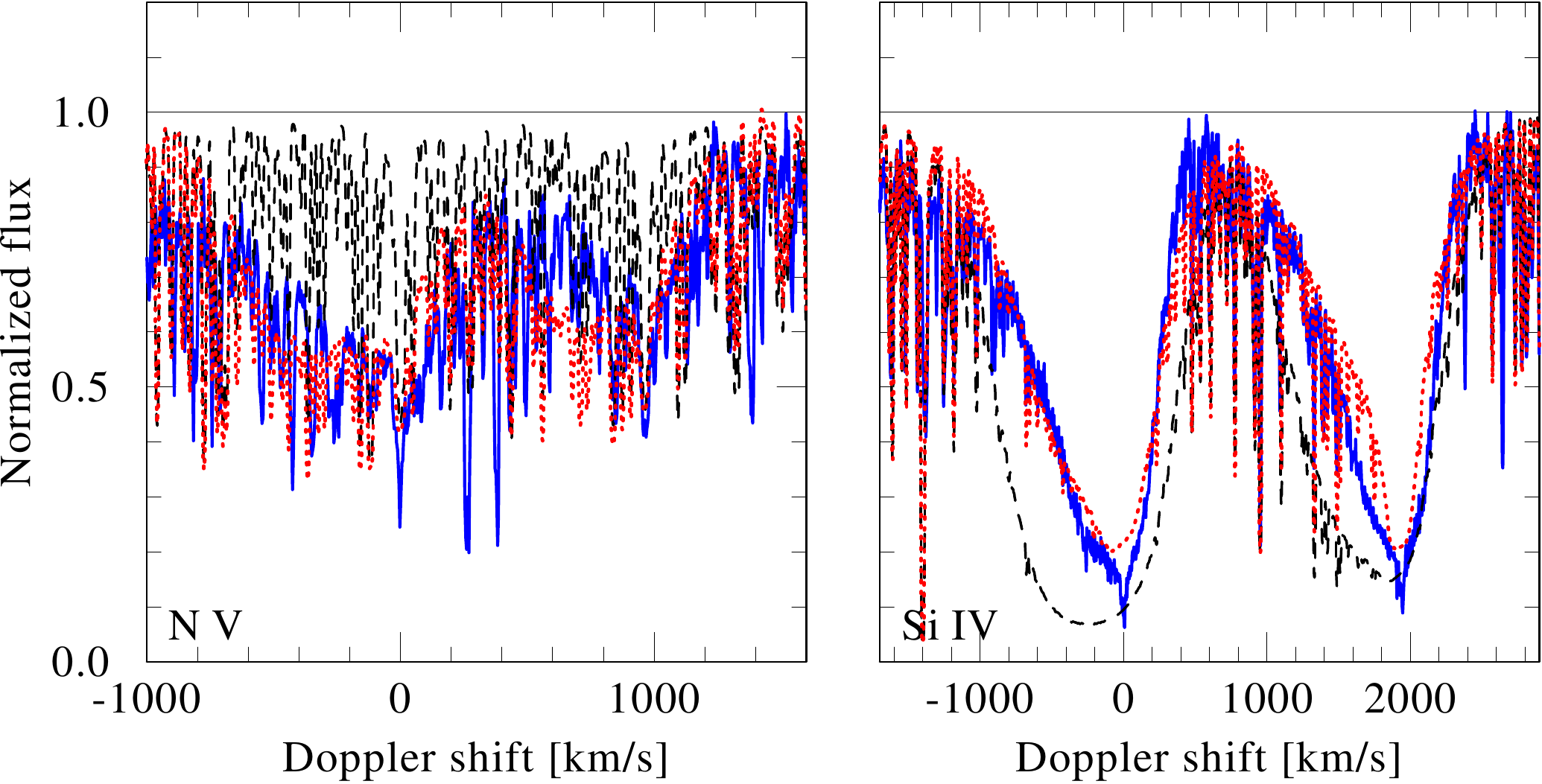}
  \caption{Normalized {\em HST} observations of {\targetname} (blue solid line) compared to our best-fitting model (red dotted line) and to the 
  same model without the inclusion of X-rays (black dashed line).}
\label{fig:HSTXrays}
\end{figure}

\section{Evolutionary status}
\label{sec:evolution}

Using the parameters given in Table\,\ref{tab:specan}, we can derive an evolutionary scenario for {\targetname}. 
Here, we use the BONNSAI\footnote{The BONNSAI web-service is available at www.astro.uni-bonn.de/stars/bonnsai} tool \citep{Schneider2014}, which implements 
Bayesian statistics on a set of evolutionary tracks for massive stars \citep{Brott2011} to constrain the best-fitting evolutionary channel. Previously 
performed by C2015, we repeat this procedure in light of the different parameters derived in this study. The parameters used are $\log L, \log T$, and $\log g_*$. 
The errors for $\log T_*$ and $\log g_*$, and $\log L$ are taken from Table\,\ref{tab:specan}. The peculiar C and N abundances are ignored.

Our BONNSAI solution predicts
an initial mass of $M_\text{ini} = 16\pm1\,M_\odot$ and an age of $5\pm1\,$Myr. Our results are consistent with 
the findings by C2015 within errors.  
We do not find any indication that a merger event was involved in the formation of {\targetname}, such as rejuvenation or rapid rotation.
Nevertheless, traces for a past merger event may be unobservable 
if it occurred several million years ago.

We note that the evolution models used by the BONNSAI tool do not include the effect of magnetic fields. 
Thus, in addition to the uncertainties involved in evolution models of massive stars 
(e.g., overshooting, mass-loss rates), systematic errors due to the omission of magnetic fields in the models 
may interfere with our results. For example, 
according to our results in Sect.\,\ref{subsec:wind}, 
{\targetname} loses significantly less mass than assumed by the BONNSAI evolution tracks. 
Because the magnetic field is known to strongly 
dampen the surface rotation, and in light of the very low $v \sin i$ value measured in this work, 
we used tracks which neglect rotationally induced mixing 
by setting the equatorial rotational velocity to zero 
($v_\text{rot, ini} = 0\,${\kms}). 

\section{X-ray spectral analysis}
\label{sec:xrays}

To analyze the {\em XMM-Newton} spectra, we used the standard
spectral fitting software {\sc xspec} \citep{Arnaud1996}. The abundances
were adopted from our spectral analysis in Sect.\,\ref{sec:specan} (cf.\ Table\,\ref{tab:abun}).  
The X-ray flux of {\targetname} in the 0.2--12\,keV band  measured by the {\em XMM-Newton} is 
$\approx 2 \times 10^{-13}$\,erg cm$^{-2}$ s$^{-1}$ (see 
Table\,\ref{tab:par_xray}). The unabsorbed flux corresponds to an X-ray 
luminosity at the distance of $1.0$\,kpc $\log{L_{\rm X}}\approx 32.0\,\text{erg}\,{\text s}^{-1}$, 
resulting in $\log{L_{\rm X}/L_{\rm bol}}\approx -6$ (cf.\ Table\,\ref{tab:specan}). This is a 
higher value compared to other stars with similar spectral 
types. For example, the X-ray luminosity and the temperature of the 
hottest plasma found in  $\mu$\,Col (O9.5~V), $\zeta$\,Oph (O9.2I~V), 10\,Lac 
(O9~V), $\sigma$\,Ori (O9.5~V), and other non-magnetic late O-dwarfs are at least an order 
of magnitude lower than in {\targetname} \citep{Oskinova2006, 
Waldron2007,Huenemoerder2012}. 

In contrast, all known magnetic O-dwarfs -- $\theta^1$\,Orionis\,C (O7~Vp), 
\object{Tr\,16-22} (O8.5~V), \object{HD\,57682} (O9.5~V), \object{$\tau\,$Sco} -- have relatively hard X-ray 
spectra and X-ray luminosities
$\msim 10^{32}$\,erg\,s$^{-1}$, similar to {\targetname} \citep{Schulz2000,Naze2014}. Thus, it 
appears safe to conclude that there is a clear dichotomy in the X-ray properties 
among non-magnetic and magnetic O-dwarfs, with the latter being 
significantly more X-ray luminous and displaying  harder X-ray spectra than the 
former (the situation may be more complex in case of OB-supergiants and Of?p 
stars, see recent reviews by \citealt{Naze2014,udD2016}). On this basis, we 
attribute the relatively high X-ray luminosity of {\targetname} to its 
magnetic nature.   
  
It order to establish the temperature of the X-ray emitting plasma, 
we  have analyzed the observed low-resolution {\em XMM-Newton} spectra of 
{\targetname}
(high-resolution RGS spectra have an insufficient signal and are not useful). The X-ray spectra 
of magnetic hot stars are usually well described by multi-temperature 
thermal plasma models \citep[e.g.,][]{Oskinova2011, Naze2014}. This is also the 
case for {\targetname}. 

As a first step, we fitted the observed spectra in the 0.2-10.0\,keV band
with a thermal two-temperature spectral model that assumes 
optically thin plasma in collisional equilibrium. At this step, the 
absorption was modeled as originating in the cold interstellar medium 
\citep{Wilms2000}.  The resulting spectral fit is statistically 
significant, with a reduced $\chi^2 =1.3$.
The best fit parameters are shown in Table~\ref{tab:par_xray}.

There is no physical motivation for restricting the plasma temperature 
distribution to two temperatures. In fact, spectroscopic analyses
show a continuous temperature  
distribution in hot stars \citep{Woj2005}. Therefore, as a next step,  
we fitted to the observed EPIC spectra  of \targetname\ a three-temperature 
spectral model. Also in this case, the fit is 
statistically significant, with $\chi^2 =1.1$. 
The three-temperature model fit is shown in Fig.\,\ref{fig:xrays}, 
and the associated parameters are shown in Table~\ref{tab:par_xray}.
The three temperature fit reveals the hottest plasma component with a 
temperature of about 2\,keV ($\sim$\,20\,MK). Better SNR 
is required to confirm the presence of this high-temperature 
plasma. We also tried to fit a four-temperature plasma model. However, the 
quality of the data is not sufficient to further constrain the 
temperature distribution.

\begin{figure}
\centering
\includegraphics[height=0.99\columnwidth, angle=-90]{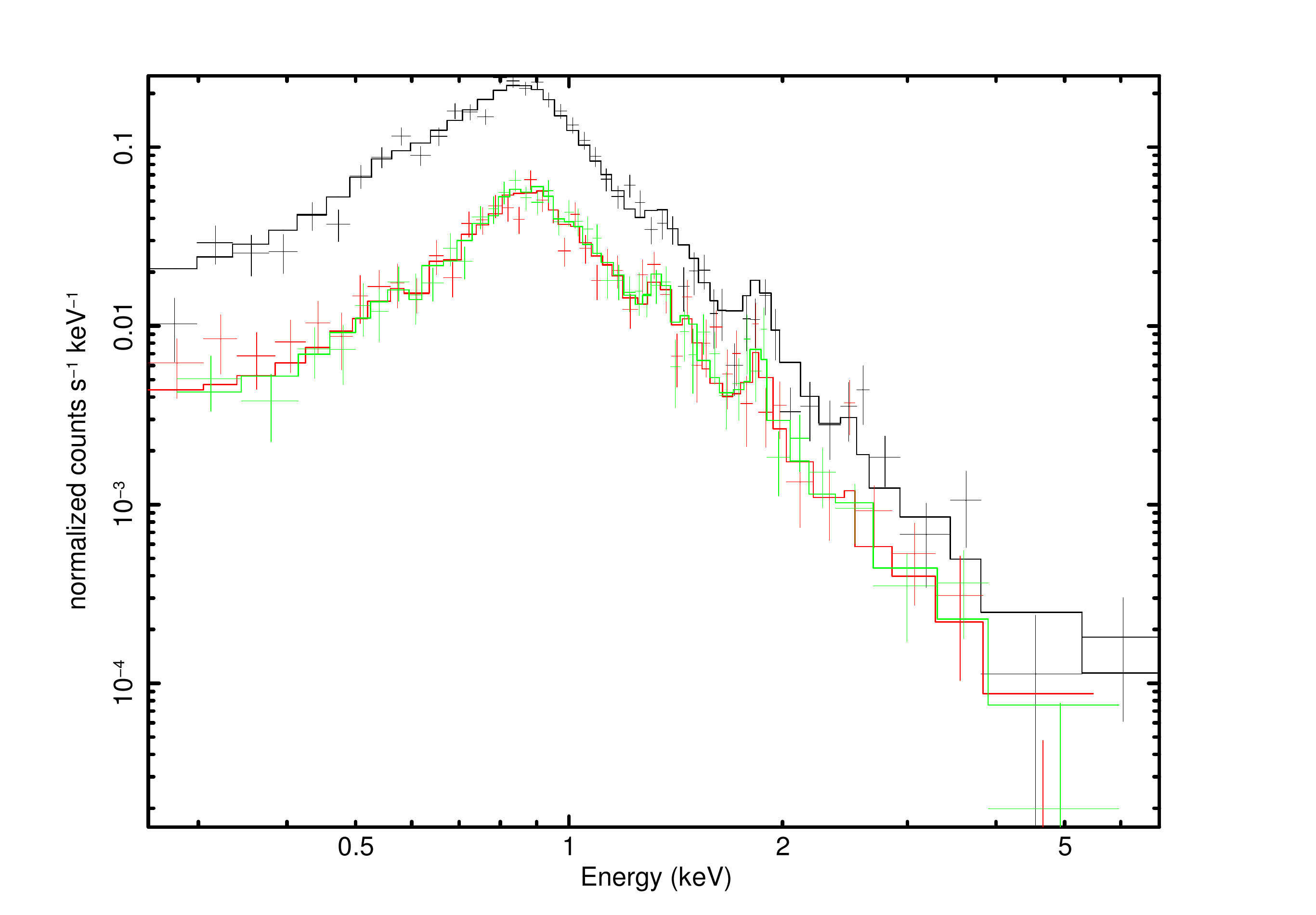}
\caption{{\em XMM-Newton} PN,  MOS1, and MOS2
spectra of {\targetname} (black, red, and green curves, respectively)  
with error bars corresponding to 3$\sigma$ with 
the best fit three-temperature thermal model (solid lines). The 
model parameters are shown in Table 1.  
}
\label{fig:xrays}
\end{figure}

\begin{table}
\begin{center}
\caption[ ]{The X-ray spectral parameters derived from the \xmm\ 
observations of {\targetname} assuming a three-temperature plasma model.}
\label{tab:par_xray}
\footnotesize
\begin{tabular}[]{lc}
\hline
\hline
\hline
\multicolumn{2}{l}{Two temperature thermal model}\\
\hline
$N_{\mathrm H}$ [$10^{21}$ cm$^{-2}$]                               
                &$4.2\pm 0.4$                    \\
                
$kT_1$ [keV]                                                                    
 
                                                 &$0.18\pm0.01$   \\
$EM_1$ [$10^{54}$ cm$^{-3}$]                                                    
                                 &$6.60\pm0.4$   \\
$kT_2$ [keV]                                                                    
 
                                                 &$0.73\pm0.02$   \\
$EM_2$ [$10^{54}$ cm$^{-3}$]                                                    
                                 &$2.46\pm0.03$   \\    
$\langle k T \rangle \equiv \sum_i k T_i \cdot EM_i / \sum_i EM_i$ [keV]        
 
      &0.33   \\                                 
\hline
\multicolumn{2}{l}{Three temperature thermal model}\\
\hline
$N_{\mathrm H}$ [$10^{21}$ cm$^{-2}$]                               
                &$3.6\pm 0.5$                    \\
$kT_1$ [keV]                                                                     
                                                 &$0.18\pm0.01$   \\
$EM_1$ [$10^{54}$ cm$^{-3}$]                                                    
                                 &$3.70\pm0.01$   \\
$kT_2$ [keV]                                                                     
                                                 &$0.74\pm0.02$   \\
$EM_2$ [$10^{54}$ cm$^{-3}$]                                                    
                                 &$2.01\pm0.05$   \\
$kT_3$ [keV]                                                                    
 
                                                 &$2.1\pm 0.6$   \\
$EM_3$ [$10^{54}$ cm$^{-3}$]                                                    
 
                                &$0.31\pm0.01$   \\                             
   
$\langle k T \rangle \equiv \sum_i k T_i \cdot EM_i / \sum_i EM_i$ [keV]         
      &0.46   \\
Flux [$10^{-13}$ erg cm$^{-2}$ s$^{-1}$]                                     
                       &1.6   \\
\hline

$L_{\mathrm X}\tablefootmark{a}$ [erg s$^{-1}$]                                      
                          &$1 \times 10^{32}$   \\
$\log L_{\mathrm X} / L_{\mathrm {bol}}$                                         
                             &$-6$   \\
\hline

\end{tabular}
\tablefoot{
\tablefoottext{a}{corrected for the interstellar absorption in the 0.2--12 keV band}
}

\end{center}
\end{table}

In the framework of the magnetically confined wind shocks (MCWS) model 
\citep{Babel1997, ud-Doula2002}, the wind plasma streams
collide at the magnetic equator and give rise to a shock that heats the
plasma. Hence, the maximum temperature follows from a Rankine-Hugoniot 
condition and cannot exceed a value determined by the maximum streaming velocity.
Motivated by the recent discovery of a non-thermal component in the X-ray emission of 
the magnetic star HR\,7355 \citep{Leto2017}, we also fitted the observed 
spectra with a composite two temperature thermal plasma and a non-thermal, power law
component. The model provides a fit of similar quality 
to the observed spectra ($\chi^2 =1.2$ for 138 degrees of 
freedom). The best fit thermal plasma temperatures are $kT_1=0.18\pm
1.2\,$[keV], $kT_2=0.74\pm 0.02\,$[keV], and the power-law exponent is $\Gamma=2.5\pm0.7$.
Thus, the quality of the data is not sufficient to discriminate between 
three-temperature thermal and two-temperature thermal plus non-thermal 
component models. 

Interestingly, the total absorption derived from our X-ray analysis is found to 
be somewhat higher than 
that derived from the analysis of UV and optical data (Sect.\,\ref{sec:specan}).
Using the averaged relation 
$N_{\rm H}=E_{B- V} \cdot 6\cdot 10^{21}$\,cm$^{-2}$ \citep{Gudennavar2012}, 
$E_{B - V} = 0.35$ implies $N_{\rm H, ISM}=2.1\times10^{21}$\,cm$^{-2}$, which is 
lower by almost a factor of two compared to what is found from the X-ray data 
(see Table\,\ref{tab:par_xray}). 
Since the absorption derived from X-rays 
includes both interstellar and intrinsic absorption, 
it is possible that the additional X-ray 
absorption occurs in the magnetosphere of {\targetname}. Alternatively, this discrepancy 
could be merely a consequence of the large error on $N_{\rm H}$ (see Table\,\ref{tab:par_xray}) when derived 
from our low resolution X-ray spectra, and of the uncertain relations  
between $N_{\rm H}$, $E_{B- V}$, and the extinction parameters $A_{V}$ and $R_{V}$ \citep[e.g.,][]{Gueber2009}.

To test whether some X-ray absorption originates in the magnetosphere,
we searched for the presence of absorption edges corresponding to 
the leading ionization stages of abundant metals in \targetname. Fitting the 
observed spectra with an absorption model that includes multiple ions did not 
statistically improve the fit. Hence, we concentrated on searching absorption 
edges of individual ions. Fitting a three-temperature plasma model that, 
in addition to the interstellar absorption, is also attenuated by a ``warm'' 
absorber resulted in a detection of an absorption edge at $0.61 \pm 
0.15$\,keV with an optical depth $0.9\pm0.6$. This energy corresponds to the 
absorption edges of O\,{\sc iii-vii} \citep{Verner1995}. The fact that these 
are leading ionization stages of oxygen in the magnetosphere of \targetname\ 
provides 
additional support for the detection of the absorption edge. 
Including ionization edges in the X-ray spectral model does not significantly affects the derived value of the total ISM absorption $N_\text{H,ISM}$.
Hence, while the data agree with some X-ray absorption due to the warm material trapped in the magnetosphere, the quality of our data 
prevents us from measuring the warm absorption component unambiguously.

In Sect.\,\ref{subsec:wind}, we showed that the true wind parameters ($\dot{M}, v_\infty$) cannot be derived 
for this object based on UV/optical spectra. However, as suggested by \citet{Lucy2012}, it is possible that a significant 
amount of matter leaves the star in a very hot phase and emits in X-rays. From high resolution X-ray spectra, one 
could study in detail the properties of the stellar wind of the {\targetname}, if it indeed exists. To shed light on 
the urgent weak wind problem, as well as on the true mass-loss rates of massive main sequence stars, we therefore encourage 
future observational campaigns to acquire high resolution X-ray spectra for {\targetname}.

\section{Variability}
\label{sec:variability}

\subsection{Photometric variability}
\label{subsec:photovariability}

In Fig.\,\ref{fig:lightcurve}, we present 
a lightcurve collected by the All Sky Automated Survey 3 (ASAS3, six-pixel wide aperture) for our target \citep{Pojmanski2002}. 
The dataset was scanned for significant periods using Fourier and F-statistics, but none 
could be identified. Striking are the sudden decreases in brightness, which only last for $\approx 1\,$d, evident 
in Fig.\,\ref{fig:lightcurve}. We checked whether these events occur periodically and can safely 
reject this possibility, excluding binary eclipses as a possible explanation. Sudden brightness changes could also 
arise from outbursts. However, the intensity of change in magnitude and the short time scale over which these events 
occur do not plausibly agree with such a model. Similar outliers can be seen 
in other lightcurves obtained by the ASAS \citep[e.g.,][]{Paczyski2006}.
We conclude that these events are most likely observational artefacts.

\begin{figure}[!htb]
\centering
  \includegraphics[width=\columnwidth]{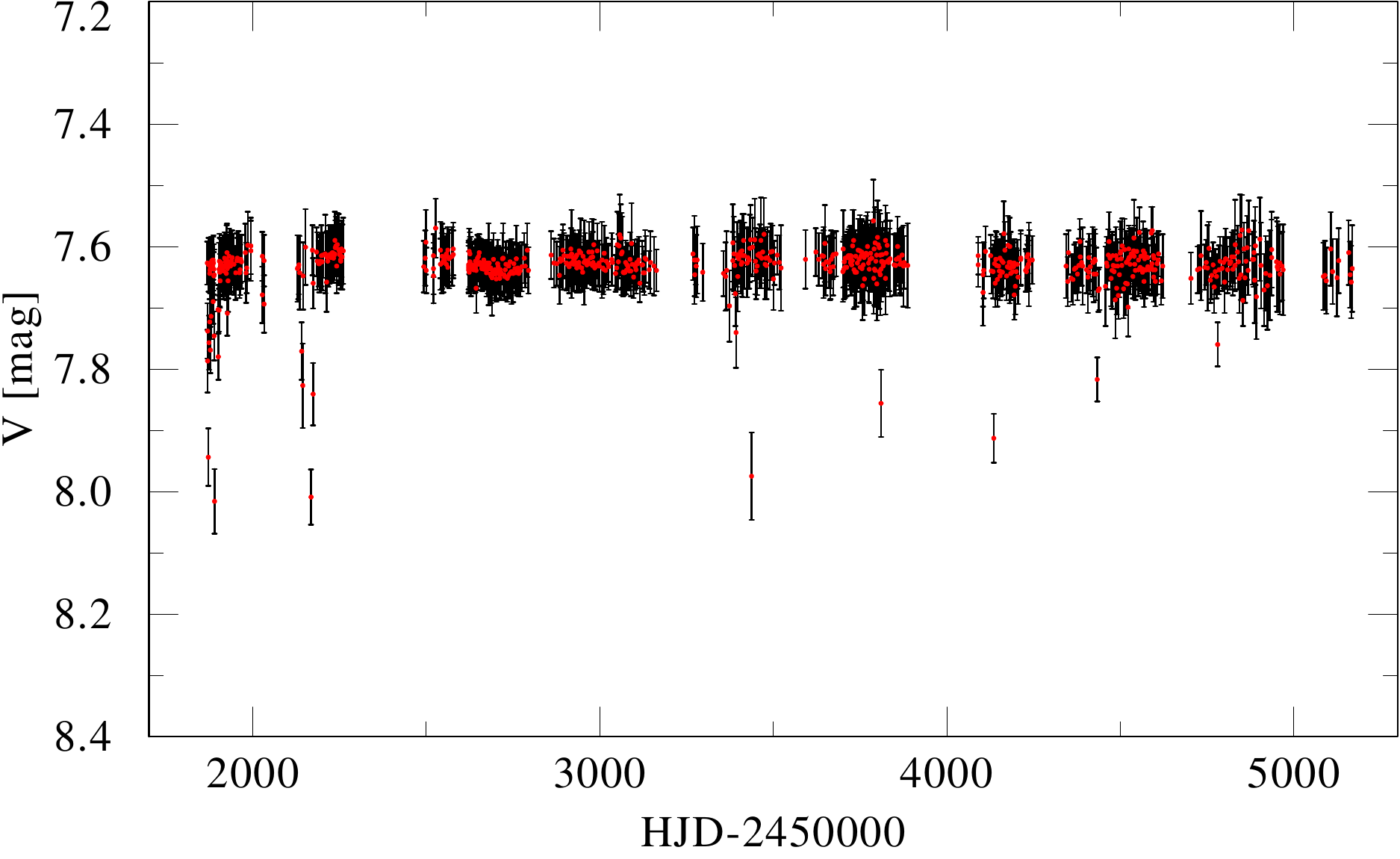}
  \caption{ASAS3 lightcurve of {\targetname}}
\label{fig:lightcurve}
\end{figure} 

\subsection{Spectroscopic variability}
\label{subsec:specvariability}

{\targetname} appears to exhibit both small-scale and large-scale variability that are especially 
evident in the H$\alpha$ line (see Fig.\,\ref{fig:Halpha_var}). 
While the dynamic structure of the magnetosphere implies some stochastic variability 
\citep[e.g.,][]{Sundqvist2012, ud-Doula2013}, 
it is also expected that a periodic variability that correlates 
with the rotational period of the star should be present \citep[e.g.,][]{Stibbs1950, Townsend2005, Wade2011}. 

\begin{figure}[!htb]
\centering
  \includegraphics[width=\columnwidth]{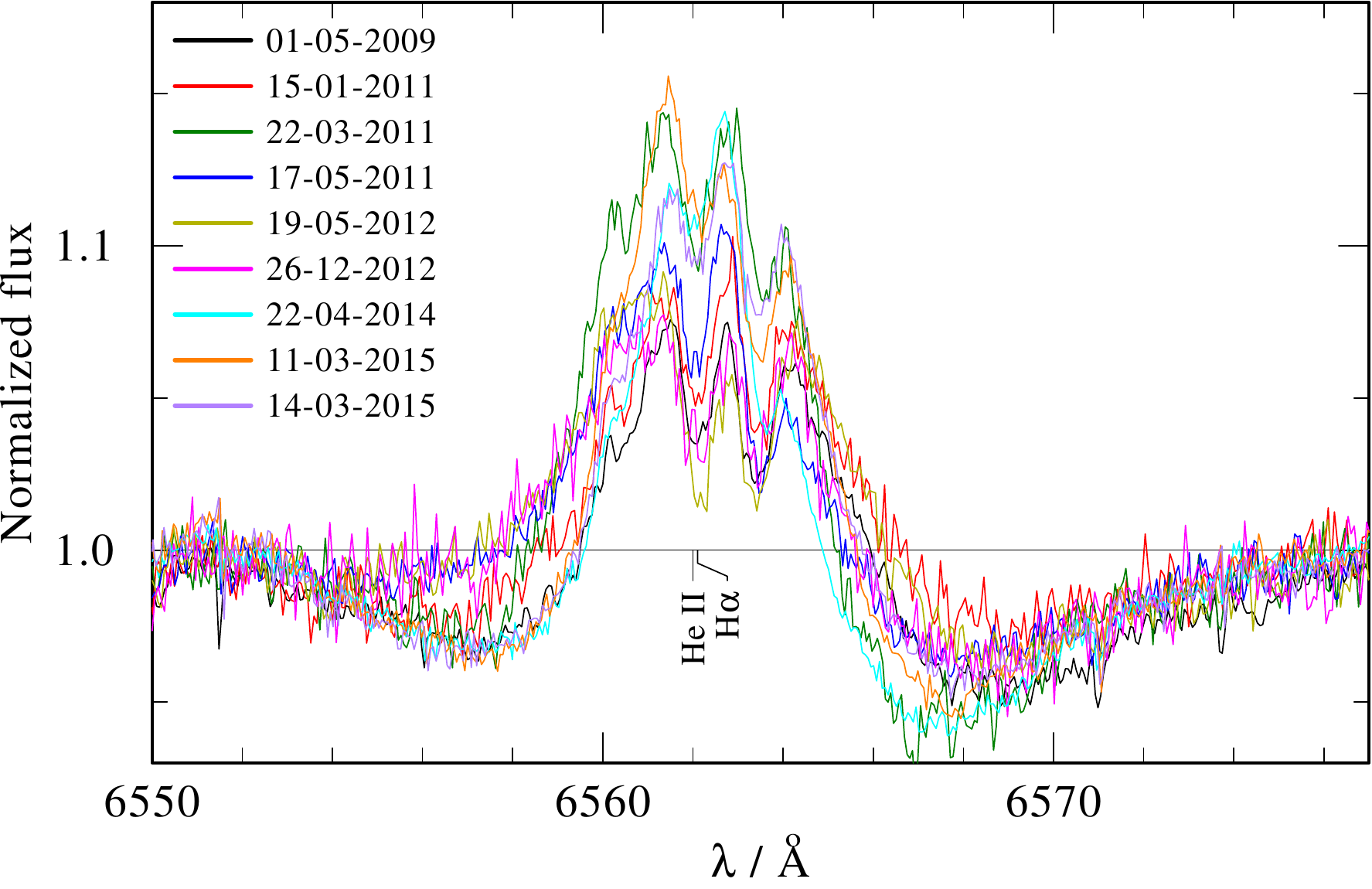}
  \caption{FEROS, FIES, and HARPS spectra of H$\alpha$ taken in the years 2009 -- 2015 (see legend and Table\,\ref{tab:obslog})}
\label{fig:Halpha_var_all}
\end{figure}

To illustrate the short scale variability, we refer the reader to Fig.\,\ref{fig:Halpha_var} in Sect.\,\ref{sec:specan}.
The figure shows that, while H$\alpha$ maintains a similar equivalent width between 11-03-2015 and 14-03-2015, 
some variability can be seen in the strength of the leftmost emission peak and the right absorption wing. In other words, the star exhibits 
some variability on the scale of days, which can be considered short-scale in our context.
The third spectrum,
taken about half a year prior to the others (April 22, 2014), shows a higher equivalent width (less emission) and a narrower profile. This suggests that 
a mechanism is present that is responsible for a long-term variability.
Fig.\,\ref{fig:Halpha_var_all} 
shows nine spectra of H$\alpha$ taken in the years 2009 -- 2015 (see Table\,\ref{tab:obslog} and legens). One can see clear changes in the equivalent 
widths and full width half maxima (FWHM) of the profiles, as well as differing strengths of the absorption wings.

These spectra alone are not sufficient to determine a period. For this reason, we acquired  35 amateur spectra (P.\ Luckas, priv.\ comm.) collected 
specifically for this project (see Sect.\,\ref{sec:obsdata}). The spectra are of lower quality, but offer the advantage of a much denser time 
coverage. A few selected spectra are shown in Fig.\,\ref{fig:Halpha_Luckas}.

\begin{figure}[!htb]
\centering
  \includegraphics[width=\columnwidth]{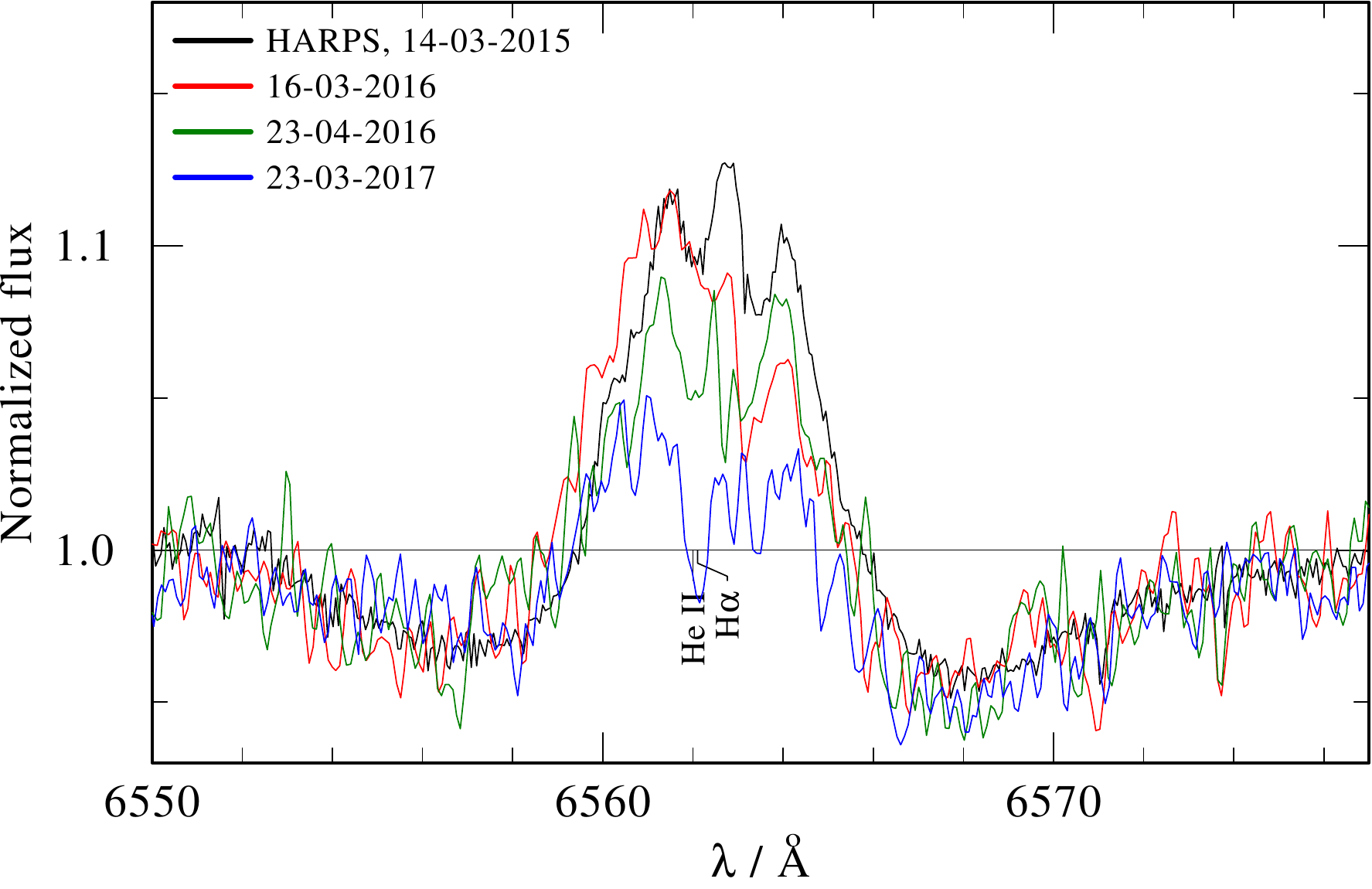}
  \caption{A few amateur SPO spectra of H$\alpha$, shown with a single HARPS spectrum for comparison}
\label{fig:Halpha_Luckas}
\end{figure}

It is not trivial to identify the quantity whose time series would best constrain a period for the global variability. A rotational modulation can 
lead to partial polar/equatorial view (relative to the magnetic axis) of the disk-like, H$\alpha$ emitting structure, which in turn 
leads to a periodic variation in the equivalent widths \citep[e.g.,][]{Sundqvist2012, ud-Doula2013}. We therefore measured the equivalent width of the H$\alpha$ line 
in the range $6551 - 6576\,\AA$ in all available spectra. The error bars are attributed mostly to uncertainty in rectification, and are 
roughly estimated by $\Delta \lambda\,/\,\text{SNR}$, where $\Delta \lambda = 25\,\AA$ is the integration domain.
The results are listed in Table\,\ref{tab:obslog} and shown in Fig.\,\ref{fig:EWstime}.

\begin{figure}[!htb]
\centering
  \includegraphics[width=\columnwidth]{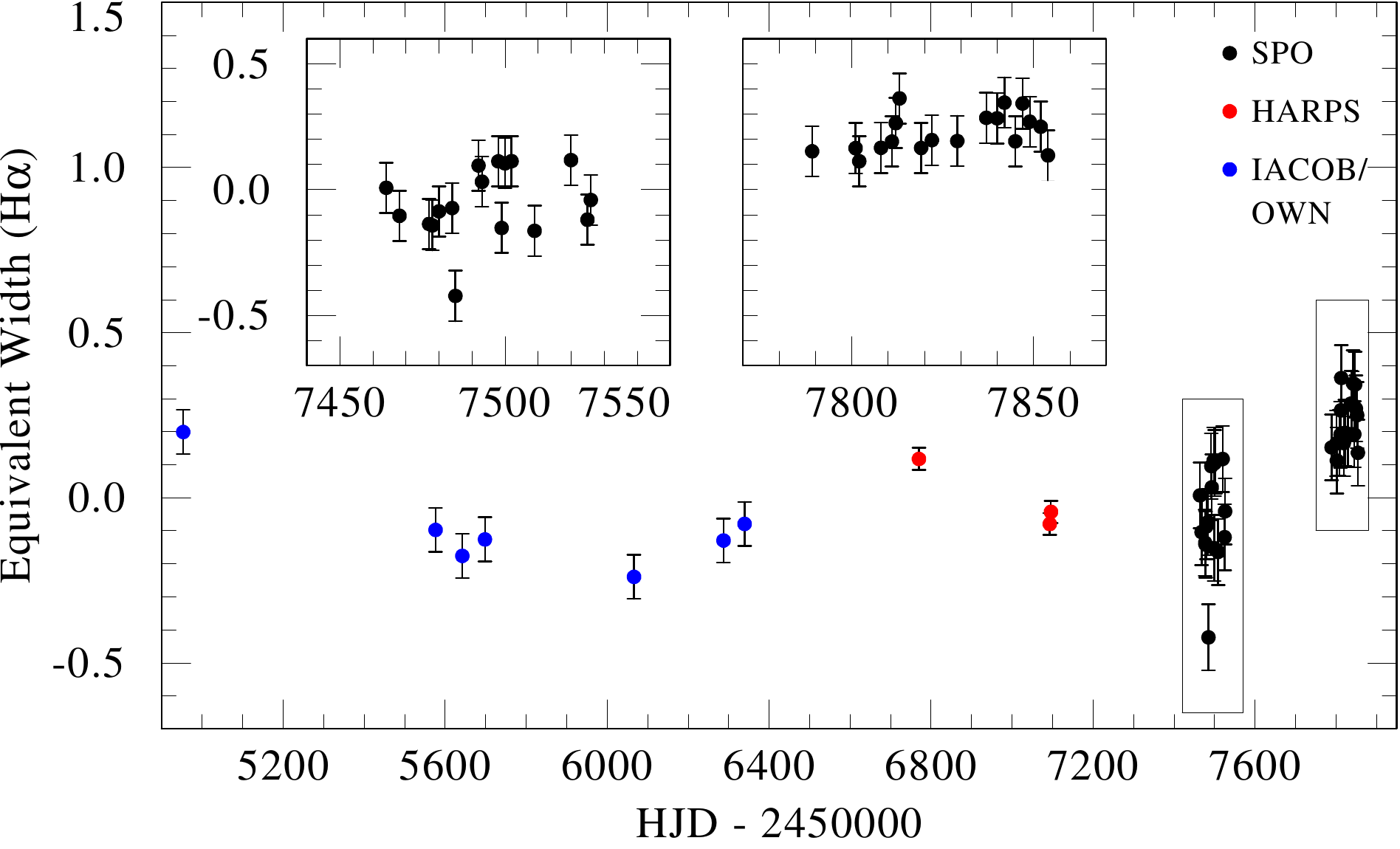}
  \caption{Equivalent widths of H$\alpha$ versus Heliocentric Julian date, as measured in the IACOB/OWN, HARPS and amateur SPO 
  spectra}
\label{fig:EWstime}
\end{figure} 

Figure\,\ref{fig:EWstime} seems to be suggestive of a modulation of the 
equivalent widths over a very long period, of the order of years. A Fourier analysis of the signal 
\citep{Scargle1982, Horne1986} suggests a period of $\approx 5\,$yr as the only significant period.
This result is consistent with the upper limit of $v \sin i < 4\,${\kms} derived 
in Sect.\,\ref{sec:specan}. From the derived period and stellar radius, and 
assuming the probable value $i \approx 60^\circ$, this suggests a very low rotational 
velocity of $v_\text{eq} \approx 0.2\,${\kms}.
Additional attempts to constrain the period using other proxies such as the line width or its centroid did not result 
in any coherent periods. We therefore suggest in this paper a rotational period of $\approx 5\,$yr, but more spectra will be 
necessary to validate our claim. Overall, this result is consistent with the fact that magnetic stars are very slow rotators that lost 
their angular momentum via magnetic braking \citep{Weber1967, ud-Doula2002}.

\section{Summary}
\label{sec:summary}

We performed a comprehensive, multiwavelength analysis of {\targetname} (O9.7 V). With a  
dipole magnetic field of $B_\text{d} \gtrsim 2\,$kG, it is one of about ten magnetized O-stars known. 
Using high quality X-ray, UV, and optical spectra acquired by the {\em XMM-Newton}, {\em HST}, and 
HARPS instruments, respectively, we could derive the X-ray properties of {\targetname} and analyze its atmosphere and wind. 
Moreover, 45 spectra were used to constrain a periodic variability of the 
H$\alpha$ line. We conclude the following:

\begin{itemize}
 \item The fundamental stellar parameters ($T_*, L, \log g$) of {\targetname} are typical for its spectral type.
       Nitrogen and carbon abundances are found to be sub-solar.
 \item The projected rotational velocity $v \sin i$, the microturbulent velocity $\xi_\text{ph}$, and the macroturbulent velocity $v_\text{mac}$, 
       are all found to be smaller than $4\,$\kms.
 \item The X-ray spectrum can be well-fitted with a thermal model accounting for either two or three components, 
       and implies a higher-than-average X-ray luminosity ($\log L_\text{X} / L_\text{bol} = -6.0$). In the three-component model,  
       the X-ray temperature reaches values up to $T_\text{X}\approx 20\,$MK.
 \item Variability of H$\alpha$ equivalent widths is suggestive of a very long period of the order of $\approx5\,$yr, consistent with the low $v \sin i$ value.
 \item The mass-loss rate that our target would have in the absence of a magnetic field could be roughly estimated to be  $\log \dot{M}_{B = 0} \approx -9.0\,${{\myr}}. This is 
       significantly less than theoretically predicted \citep{Vink2000}, but is in line with what is found for non-magnetic stars of similar spectral types \citep{Marcolino2009}. 
       With an Alfv\'en radius of $r_\text{A} \gtrsim 12\, R_*$, the true mass-loss rate of {\targetname} is estimated as $\log \dot{M} \lesssim -10.2$\,{{\myr}}.      
\end{itemize}

To conclude, we would like to point out that, as our study illustrates, 
slowly rotating magnetic stars can provide important constraints on the  \emph{weak wind problem}.
The spectra of main sequence OB-type stars often exhibit very little or no signatures for a stellar wind in UV spectra, which contain the 
principle diagnostics for mass-loss rates. Magnetic fields that confine the stellar winds enhance the circumstellar densities and result in clear 
spectral features. Using sophisticated models for the magnetospheres of massive magnetic stars, one can therefore infer values for 
$\log \dot{M}_{B = 0}$. These should be similar to \emph{true} mass-loss rates for \emph{non-magnetic} main sequence stars of the same spectral type.

\begin{acknowledgements}
T.S. and L.O. acknowledge support from the german "Verbund-
forschung" (DLR) grants 50 OR 1612 and 50 OR 1302.
A.S. is supported by the Deutsche Forschungsgemeinschaft (DFG) under grant HA 1455/26.
The IACOB spectroscopic database is based on observations made with the Nordic Optical Telescope [www.not.iac.es] operated by
the Nordic Optical Telescope Scientific Association, and the Mercator Telescope [www.mercator.iac.es], operated by the Flemish
Community, both at the Observatorio de El Roque de los Muchachos [www.iac.es] (La Palma, Spain) of the Instituto de Astrofísica
de Canarias [www.iac.es]. This research has made use of the VizieR catalogue access tool, CDS,
Strasbourg, France. The original description of the VizieR service was
published in A\&AS 143, 23. We thank our referee, G.\ Wade, for his
  constructive and careful reviewing of our manuscript.
\end{acknowledgements}
\bibliography{literature}

\Online

\begin{appendix}
\section{Observations and H$\alpha$ equivalent widths}
\label{sec:app}

\renewcommand{\arraystretch}{1.1}

\begin{table}[!htb]
\scriptsize
\small
\caption{Compilation of optical observations and measured H$\alpha$ equivalent widths for {\targetname}.}
\label{tab:obslog}
\begin{center}
\begin{tabular}{c l c c r}
\hline
\rule{0pt}{2.5ex}    
No.& Instrument &  Date       &      HJD   & ${\rm EW}_{{\rm H}\alpha}$ [$\AA$] \\
\hline      
\rule{0pt}{2.5ex}    
1  & FEROS  &  01-05-2009     & 2454953.54 & -0.11 \\
2  & FIES   &  15-01-2011     & 2455576.54 & -0.17 \\
3  & FEROS  &  22-03-2011     & 2455642.54 & -0.45 \\
4  & FEROS  &  17-05-2011     & 2455698.54 & -0.27 \\
5  & FEROS  &  19-05-2012     & 2456066.54 & -0.46 \\
6  & FIES   &  26-12-2012     & 2456287.54 & -0.17 \\
7  & FIES   &  16-02-2013     & 2456339.54 & -0.12 \\
8  & HARPS  &  22-04-2014     & 2456770.04 &  0.02 \\
9  & HARPS  &  11-03-2015     & 2457093.04 & -0.09 \\
10 & HARPS  &  14-03-2015     & 2457096.04 & -0.18 \\
11 & SPO &  16-03-2016     & 2457464.02 &  0.00 \\
12 & SPO &  20-03-2016     & 2457468.02 & -0.10 \\
13 & SPO &  29-03-2016     & 2457476.99 & -0.13 \\
14 & SPO &  30-03-2016     & 2457477.98 & -0.14 \\
15 & SPO &  01-04-2016     & 2457479.97 & -0.08 \\
16 & SPO &  05-04-2016     & 2457483.98 & -0.07 \\
17 & SPO &  06-04-2016     & 2457484.97 & -0.42 \\
18 & SPO &  13-04-2016     & 2457491.96 & 0.09 \\
19 & SPO &  14-04-2016     & 2457493.01 & 0.03 \\
20 & SPO &  19-04-2016     & 2457497.96 & 0.11 \\
21 & SPO &  20-04-2016     & 2457498.97 & -0.15 \\
22 & SPO &  21-04-2016     & 2457499.97 & 0.10 \\
23 & SPO &  23-04-2016     & 2457501.97 & 0.11 \\
24 & SPO &  30-04-2016     & 2457508.94 & -0.16 \\
25 & SPO &  11-05-2016     & 2457519.96 & 0.11 \\
26 & SPO &  16-05-2016     & 2457524.93 & -0.11 \\
27 & SPO &  17-05-2016     & 2457525.94 & -0.04 \\
28 & SPO &  04-02-2017     & 2457789.07 & 0.15\\
29 & SPO &  16-02-2017     & 2457801.03 & 0.16\\
30 & SPO &  17-02-2017     & 2457802.03 & 0.11\\
31 & SPO &  23-02-2017     & 2457808.02 & 0.16 \\
32 & SPO &  26-02-2017     & 2457811.03 & 0.19 \\
33 & SPO &  27-02-2017     & 2457812.09 & 0.26 \\
34 & SPO &  28-02-2017     & 2457813.11 & 0.36 \\
35 & SPO &  06-03-2017     & 2457819.07 & 0.16 \\
36 & SPO &  09-03-2017     & 2457822.02 & 0.19 \\
37 & SPO &  16-03-2017     & 2457829.02 & 0.19 \\
38 & SPO &  24-03-2017     & 2457837.00 & 0.28 \\
39 & SPO &  27-03-2017     & 2457839.99 & 0.28 \\
40 & SPO &  29-03-2017     & 2457841.99 & 0.34 \\
41 & SPO &  01-04-2017     & 2457844.99 & 0.19 \\
42 & SPO &  03-04-2017     & 2457846.98 & 0.34 \\
43 & SPO &  05-04-2017     & 2457848.98 & 0.26 \\
44 & SPO &  08-04-2017     & 2457851.97 & 0.24 \\
45 & SPO &  10-04-2017     & 2457853.97 & 0.13 \\
\hline
\end{tabular}
\end{center}
\end{table}

\end{appendix}

\end{document}